\documentclass[superscriptaddress, aps, prl, preprint,  amsmath, showkeys, a4paper]{revtex4-1}
\usepackage{subfig, multirow, rotating,xcolor}
\usepackage[]{graphicx}
\usepackage[section]{placeins}
\usepackage[printfigures]{figcaps}
\figcapsoff 
\usepackage[font=small,justification=justified,singlelinecheck=false]{caption}
\usepackage{mathrsfs}
\usepackage{xr-hyper}
\usepackage{hyperref}
\usepackage{cleveref}
\usepackage{accents}
\usepackage{filecontents}
\renewcommand\eqref[1]{Eq.~(\ref{#1})}
\newcommand\figref[1]{Figure~\ref{#1}}

\newcommand{\bc}{b_{q}^{\dagger}}

\newcommand{\ba}{b_{q}}

\newcommand{\ene}{\varepsilon}

\newcommand{\lb}{\left(}
\newcommand{\rb}{\right)}
\newcommand{\lsb}{\left[}
\newcommand{\rsb}{\right]}
\newcommand{\lcb}{\left\{}
\newcommand{\rcb}{\right\}}
\newcommand{\Dm}{\mathcal{D}}

\begin{document}
\title[]{Dynamics with Simultaneous Dissipations to Fermionic and Bosonic Reservoirs}
\author{Elvis F. Arguelles}
\affiliation{Institute for Solid State Physics, The University of Tokyo}
\author{Osamu Sugino}
\affiliation{Institute for Solid State Physics, The University of Tokyo}

\date{\today}
\begin{abstract}
We introduce a non-phenomenological framework based on the influence functional 
method to incorporate simultaneous interactions of particles with fermionic 
and bosonic thermal reservoirs. In the slow-motion limit, the electronic friction kernel 
becomes Markovian, enabling an analytical expression for the friction coefficient.
The framework is applied to a prototypical electrochemical system, 
where the metal electrode and solvent act as fermionic and bosonic reservoirs, respectively. 
We investigate quantum vibrational relaxation of hydrogen on metal surfaces, 
showing that dissipation to electron-hole pairs reduces the relaxation time. 
Additionally, in solvated proton discharge, electronic friction prolongs charge transfer 
by delaying proton transitions between potential wells.
This study provides new insights into the interplay of solvent and electronic dissipation effects,
with direct relevance to electrochemical processes and other systems involving multiple thermal 
reservoirs.
\end{abstract}
\maketitle
\section{Introduction}

Open systems evolve over time interacting with fermionic or bosonic reservoirs 
representing electrons in electrodes, lattice vibrations or electromagnetic fields. 
In electrochemical systems, the dynamics occurs under a moderate non-equilibrium condition 
influenced by a delicate balance of the multiple reservoirs. Particularly important 
property to characterize the systems is the rate of exchanging the energy and charge 
with reservoirs, which can be investigated in principle using a microscopic 
theory-like based on the Langevin equation\cite{Langevin1908,Lemons1997,Ford1987,Ford1988} 
by incorporating the fluctuations and dissipation due to environments. 
The multiple reservoir effect, however, has not received much attention.

Over the years, the Langevin equation has been formally derived\cite{Ford1965,Schmid1982} and 
re-derived\cite{Caldeira1983,Metiu1984,Newns1985,Sebastian1987,Ford1987,Ford1988}
in various contexts, accounting for the dissipation of the impinging particle’s kinetic energy 
not only into substrate phonon excitations via phononic friction\cite{Newns1985}, 
but also into the generation of coherent electron-hole (e-h) pairs 
via electronic friction\cite{Brako1980,Schonhammer1980,Brandbyge1995,HGT1995,Plihal1998,Dou2017,Martinazzo2022}. 
The equation has subsequently been generalized for the former\cite{Newns1985,Sebastian1987} 
and the latter\cite{Hedegard1987,Brandbyge1995}, 
but their simultaneous operation has mostly been neglected, often justified by arguments based 
on the disparity of particle masses. This is, however, not always the case because the strength 
of the former is considerably dependent on the environments; 
it is reduced, for example, when the reaction center moves away from the electrode and correspondingly 
electrons must transfer farther away therefrom, as typically occurs with increasing pH\cite{Kumeda2023,Li2022}.
The crossover of the dominant dissipation channels thus occurs in nature.

To address this issue, we reformulate the particle dynamics using the influence functional 
path integral\cite{Feynman2000} framework, which yields without phenomenological assumptions a quasiclassical 
Langevin equation that incorporates non-Markovian effects of both reservoirs. 
The dissipation kernel and fluctuations become Markovian in the limit of slow particle motion,
allowing analytical evaluation of the electronic friction coefficient. 
We demonstrate this through numerical studies of (1) quantum vibrational relaxation of hydrogen on metal surfaces 
and (2) solvated proton discharge dynamics on metal electrodes. 
Considering the generality of this scheme, it should extend the frontier of the research on quasi-equilibrium systems.

\section{Theory}

The total Hamiltonian of a system comprising a particle coupled to two thermal reservoirs is expressed as 
$H = H_{p} + H_{\mathcal{R}} + H_{p\mathcal{R}}$, where 
$H_{p}$ is the particle Hamiltonian, 
$H_{\mathcal{R}}=H_{1}+H_{2}+H_{12}$ represents the composite reservoirs
($\mathcal{R}=\mathcal{R}_{1}\cup \mathcal{R}_{2}$)
with $H_{12}$ accounting for inter-reservoir interactions, and 
$H_{p\mathcal{R}}$ describes particle-reservoir coupling. 
The formalism is initially developed using a bosonic representation, 
with the case of a fermionic reservoir recovered subsequently.
%
We represent $\mathcal{R}$ as interacting sets of harmonic oscillators
$\mathcal{R}_{1}=\lcb \Omega_{1j}\mid j\in J\rcb$ and $\mathcal{R}_{2}=\lcb \Omega_{2q}\mid q\in Q\rcb$
where $\Omega_{\alpha=1,2}$ is the frequency, and $J$ and $Q$ are sets of modes
of $\mathcal{R}_{1}$ and $\mathcal{R}_{2}$, respectively. 
We describe the $\mathcal{R}_{1}$ and $\mathcal{R}_{2}$ interaction by the
mode-resolved coupling $\beta_{jq}$. Eliminating $H_{12}$ 
(Sec.~\ref{sup:tot_Ham} of Supplementary Material\cite{supp})
results in 
\begin{equation}
	H_{\mathcal{R}}=\sum_{j}\Omega_{1j}B_{j}^{\dagger}B_{j}+\sum_{q}\Omega_{2q}\bc\ba.
	\label{eq:H_R}
\end{equation}
Here, $B_{j}(B_{j}^{\dagger})$ and $\ba(\bc)$ are the boson annihilation (creation) operators 
of $\mathcal{R}_{1}$ and $\mathcal{R}_{2}$, respectively.
The eigenvalue equation, $H_{\mathcal{R}}|\varphi\rangle = \ene_{\varphi} |\varphi \rangle$ 
forms the basis of our functional integral scheme.
We factorize $|\varphi\rangle$ as $|\varphi\rangle = |n\rangle|m\rangle$,
where $|n\rangle$ and $|m\rangle$ are the eigenstates of
$\mathcal{R}_{1}$ and $\mathcal{R}_{2}$, respectively.
Denoting the particle coupling with $\mathcal{R}_{1}$ and $\mathcal{R}_{2}$ 
as $c_{j}(x)$ and $f_{q}(x)$, respectively, with $x$ being the particle coordinate, 
the total Hamiltonian then reads
\begin{equation}
	H = H_{\mathcal{R}} + \frac{p^{2}}{2m} + V(x) \
	+\sum_{j}\lsb c_{j}(x)B_{j} + c_{j}^{*}(x)B_{j}^{\dagger} \rsb\
	+\sum_{q}\lsb f_{q}(x)\ba + f_{q}^{*}(x)\bc \rsb,
	\label{eq:H_tot_1}
\end{equation}
where the second and third terms describe the particle dynamics.
At the initial time $t=t_{i}$ we assume the reservoirs are in thermal equilibrium 
and the subsystems are noninteracting. This state is described by
\begin{equation*}
	\rho(t_{i}) =\rho_{p}(t_{i})\otimes\frac{1}{Z}e^{-\beta H_{\mathcal{R}}}
\end{equation*}
where $\rho_{p}(t_{i})$ denotes the particle's initial density matrix, and
$Z=\sum_{\varphi}e^{-\beta \ene_{\varphi}}$ is the partition function. 
Rapid subsystem mixing at $t>t_{i}$ justifies neglecting initial correlations\cite{Brandbyge1995}.
The total density operator 
\begin{equation*}
	\rho(t_{f},t_{i}) =U(t_{f},t_{i})\rho(t_{i})U^{\dagger}(t_{f},t_{i})
\end{equation*}
governs the system's evolution from $t_{i}$ to $t_{f}$, where 
$U(t_{f},t_{i})=\exp\lsb-iH(t_{f}-t_{i}) \rsb$ and $U^{\dagger}(t_{f},t_{i})$ its 
corresponding adjoint. We express $\rho(t_{f})$ as a reduced density matrix defined as
\begin{equation}
	\begin{aligned}
		\rho_{red}(t_{f},t_{i}) &= \
	\int dx_{i}\int dy_{i} \rho_{p}(x_{i},y_{i},t_{i})
\Big< U(x;t_{f},t_{i}) U^{\dagger}(y;t_{f},t_{i})\Big> ,
\end{aligned}
	\label{eq:rho_t}
\end{equation}
where $\langle\ldots\rangle=\frac{1}{Z}\mathrm{Tr}[e^{-\beta H_{\mathcal{R}}}\dots]$ 
is thermal average  
over the complete sets of bosonic states in \eqref{eq:H_R}, 
$\rho_{p}(x_{i},y_{i},t_{i})\equiv\langle x_{i}|\rho_{p}(t_{i})|y_{i}\rangle$, 
$U(x;t_{f},t_{i})\equiv\langle x_{f} |U(t_{f},t_{i}) |x_{i}\rangle$ and
$U^{\dagger}(y;t_{f},t_{i})\equiv\langle y_{f} |U^{\dagger}(t_{f},t_{i}) |y_{i}\rangle$.
$x(t)$ and $y(t)$ are the forward and backward moving paths in a contour
that defines the time evolution of $U(x;t_{f},t_{i})U^{\dagger}(y;t_{f},t_{i})$.
Expressing these operators in Feynman path integral representation\cite{Feynman2005}
gives
$\rho_{red}(t_{f},t_{i})=\int dx_{i}\int dy_{i} \mathcal{J}(x_{f},y_{f},t_{f},x_{i},y_{i},t_{i}) 
\rho_{p}(x_{i},y_{i},t_{i},)$
where the transition amplitude $\mathcal{J}(x_{f},y_{f},t_{f},x_{i},y_{i},t_{i})\equiv\mathcal{J}$ 
is given by
\begin{equation}
	\mathcal{J}=\int\Dm x \int\Dm y\exp\left\{-i[S(x)-S(y)]\right\}\
	\mathscr{F}_{1}[x,y]\mathscr{F}_{2}[x,y],
		\label{eq:J}
\end{equation}
with the particle action $S(x) =\int dt \left[\frac{1}{2}m\dot{x}^{2} - V(x)\right]$.
The influence functionals $\mathscr{F}_{\alpha}[x,y]$ ($\alpha=1,2$) are 
expressed (interaction picture) as
\begin{equation}
	\begin{aligned}
		\mathscr{F}_{\alpha}[x,y] &=\Bigg< \tilde{\mathcal{T}}\exp\left[i\int dt H_{I,\alpha}(x(t)) \right]
		\mathcal{T}\exp\left[-i\int ds H_{I,\alpha}(y(s)) \right]\Bigg>,
\end{aligned}
	\label{eq:F12_xy}
\end{equation}
where, $\mathcal{T}(\tilde{\mathcal{T}})$ denotes the time (anti-time) ordering operator and  
\begin{equation*}
	\begin{aligned}
		H_{I,1}(x(t))&=\sum_{j}\lsb c_{j}(x(t))e^{-i\Omega_{1j}t}B_{j}+c_{j}^{*}(x(t))e^{i\Omega_{1j}t}B_{j}^{\dagger}\rsb \\
		H_{I,2}(x(t))&=\sum_{q}\lsb f_{q}(x(t))e^{-i\Omega_{2q}t}\ba+f_{q}^{*}(x(t))e^{i\Omega_{2q}t}\bc\rsb.
\end{aligned}
\end{equation*}
The bosonic trace is worked out in Sec.~\ref{sup:bos_trace} of \cite{supp}.
We notice that the product of the influence functionals in \eqref{eq:J} suggests that the 
reservoirs are essentially noninteracting\cite{Feynman2005}. 
We work with the influence phase $\Phi=-i\ln[\mathscr{F}]$ (\eqref{eq:inf_phase})
instead of $\mathscr{F}$ and introduce a transformation 
in terms of the average $R=\frac{1}{2}(x+y)$ and relative $r=x-y$ coordinates.
This is formally parallel to the generating functional method\cite{Schmid1982}
where $R(t)$ and $r(t)$ are the $classical$ and $quantum$ fields obtained from Keldysh rotations of
$x(t)$ and $y(t)$ residing on the forward and backward branches of the Keldysh contour\cite{Kamenev2011}.
Applying this transformation which has a Jacobian of unity modifies the 
position variables in \eqref{eq:J} as $x\to R+r/2$ and $y\to R-r/2$,
with their time-dependence implied. 
The potentials contained in $S[R,r]$ are not always
conveniently defined in this transformation\cite{Schmid1982,Newns1985} requiring a functional Taylor
expansion in $r(t)$ terminated at some order. The expansion gives 
\begin{equation}
	S(R,r) =-\int dt \left[m\ddot{R}(t)r(t) + V'(R(t))r(t)\right].
	\label{eq:S_prex}
\end{equation}
where we neglected the terms $\geq\mathcal{O}(r^{3})$ that contain 
the quantum effects of the wavepacket spreading\cite{Newns1985}.
For harmonic $V(R)$, the higher derivatives are identically zero
and $S(R,r)$ is undoubtedly quantum. For general $V(R)$, ignoring these terms
constitutes the quasiclassical approximation\cite{Schmid1982}.
We now split the influence phase as $\Phi_{\alpha} = \Pi_{\alpha} + i\Sigma_{\alpha}$,
where $\Pi\equiv\mathrm{Re}[\Phi]$ and $\Sigma\equiv\mathrm{Im}[\Phi]$ 
whose functional expansions are given in Sec.~\ref{supp:fte} of \cite{supp}. 
Consequently, the transition amplitude may then be expressed as
\begin{equation}
	\mathcal{J}=\int\Dm R \int\Dm r\exp\lcb -i\lsb S(R,r) -\sum_{\alpha}\Pi_{\alpha}^{(1)}(R,r)\rsb\rcb \
	\exp\lsb\sum_{\alpha}\Sigma^{(2)}_{\alpha}(R,r)\rsb,
\label{eq:J1}
\end{equation}
where the superscript $(n)$ denotes the order of expansion.
In \eqref{eq:Sigma2} one can deduce that $\Sigma^{(2)}\propto\frac{\delta^{2} \Sigma[R,r]}{\delta r(t)\delta r(s)}$ 
provides the stochastic fluctuations to the dynamics via the random force autocorrelation function $K(t-s)$ defined as
\begin{equation}
	K_{1}(t-s)=\frac{1}{2}\sum_{j}c'_{j}(R(t))c_{j}^{\prime *}(R(s))\
	\mathrm{coth}\lb\frac{\Omega_{1j}}{2k_{B}T}\rb\cos\Omega_{1j}(t-s).
\label{eq:force_cor}
\end{equation}
Similarly, one can show [\eqref{eq:Pi1}] that $\Pi_{1}^{(1)}\propto\frac{\delta \Pi[R,r]}{\delta r(t)}$ contains
the dissipation 
through the nonlocal friction coefficient kernel
\begin{equation}
	\Gamma_{1}(t-s)=\frac{1}{2}\sum_{j}\frac{c'_{j}(R(t))c_{j}^{\prime *}(R(s))}{\Omega_{1j}}\
	[\mathrm{sgn}(t-s)+1]\cos\Omega_{1j}(t-s).
\label{eq:fric_ker}
\end{equation}
and a potential shift 
$\Delta\Omega_{1}\equiv\sum_{j}\frac{|c_{j}(R)|^{2}}{2\Omega_{1j}}$ that renormalizes $V(R)$.
Except for the presence of a sign function, the form of $\Gamma(t-s)$ above closely 
resembles the friction kernels previously obtained by Head-Gordon and Tully\cite{HGT1995}.
Expressions for $K_{2}(t-s)$ and $\Gamma_{2}(t-s)$
are retrieved by replacing in \eqref{eq:fric_ker} and \eqref{eq:force_cor} 
their corresponding variables. The transition amplitude now takes the form 
\begin{equation}
	\begin{aligned}
		\mathcal{J}&=\int\Dm R \int\Dm r\exp\lcb i\int dt\lsb m\ddot{R}(t) + \tilde{V}'(R(t)) \
+\int ds\sum_{\alpha}\Gamma_{\alpha}(t-s)\dot{R}(s)\rsb r(t) \right.\\
&\left.-\frac{1}{2}\int dt \int ds r(t)\sum_{\alpha}K_{\alpha}(t-s)r(s)\rcb,
\end{aligned}
\label{eq:J2}
\end{equation}
where
$\tilde{V}(R)=V(R)-\sum_{\alpha}\Delta\Omega_{\alpha}$.
The summations over $\alpha$ in \eqref{eq:J2} implies the existence of an
effective random force $\xi(t)$ that accounts for
the stochastic fluctuations coming from both reservoirs.
$\xi(t)$ has the properties
$\langle\xi(t)\rangle=0$ and $\langle\xi(t)\xi(s)\rangle=\sum_{\alpha}K_{\alpha}(t-s)$.
After stochastic averaging of \eqref{eq:J2} over $\xi(t)$, 
the $r(t)$ path integral can be done explicitly (Sec.\ref{supp:stoav} of \cite{supp}).
This yields a Dirac delta functional (\eqref{eq:supp_J}) where
we deduce the generalized quasiclassical Langevin equation 
\begin{equation}
	\begin{aligned}
	m\ddot{R}(t) + \tilde{V}'(R(t)) \
	+\int ds\sum_{\alpha}\Gamma_{\alpha}(t-s)\dot{R}(s) =\xi(t)
\end{aligned}
\label{eq:Langevin_1}
\end{equation}
describing the motion of the particle interacting
simultaneously with two thermal reservoirs. 

\section{Interactions with Fermionic and Bosonic Environments}

Let us now turn to the case where one of the reservoirs is fermionic.
Specifically, we consider a system consisting of 
a particle interacting with a metal surface immersed in a solvent.
The solvent modes and the electronic manifold of the metal represent
the bosonic ($\mathcal{R}_{1}$) and fermionic ($\mathcal{R}_{2}$)
reservoirs, respectively. 
The frequency and eigenstates are renormalized by the inter-reservoir interaction,
but for simplicity, we assume that the simulation parameters have been already renormalized. 
We also assume that the particles couple linearly with the bosonic bath, 
thereby yielding the featureless bosonic friction coefficient $\gamma$ 
in the Caldeira-Leggett framework\cite{Caldeira1983_PI,Caldeira1983}
(\eqref{eq:fric_kerR1_2} of \cite{supp}).
The dynamics of solvated particles on a metal is typically described by 
the time-dependent Newns-Anderson-Schmickler (TD-NAS) model\cite{Schmickler1986,Arguelles2024}. 
In this model, the electronic coupling is treated in the wide-band limit in terms of 
$\Delta(t)=\pi\sum_{k}|V_{ak}(t)|^{2}\delta(\ene-\ene_{k})$,
with $V_{ak}(t)$ being the time-dependent hopping integrals 
and $\ene_{k}$ is the band energy. The particle level is 
renormalized as $\tilde{\ene}_{a}(t) = \ene_{a}(t) + (2Z-1)\lambda$ by
the solvent reorganization energy $\lambda$ ($Z$ is the particle charge).
We note that the TD-NAS model reduces
to the usual time-dependent Anderson-Holstein (AH) Hamiltonian\cite{Anderson1961,Holstein1959}
for $Z=0$ when $\lambda$ is re-interpreted as an average 
electron-phonon coupling.
Using the TD-NAS model, we obtained an effective
Hamiltonian\cite{Brako1981,Brako1989,Arguelles2024}
for the metal electron system perturbed by the slowly approaching
particle.
We then mapped\cite{Tomonaga1950,Brako1989,Mahan2000} 
it to $\mathcal{R}_{2}$ to obtain $f_{q}$ in \eqref{eq:H_tot_1}.
This effective Hamiltonian is given by \eqref{eq:Heh1} in \cite{supp},
where (after bosonization) we identified 
$f_{q}(R)=\lsb\frac{2\omega_{q}}{\pi^{2}\rho(\ene_{F})}\rsb^{1/2}\int dt \dot{\delta}(\ene_{F},R)\exp(i\omega_{q}t)$.
Inserting $f_{q}(R(t))$ into the $\mathcal{R}_{2}$ version of \eqref{eq:fric_ker},
yields the electronic frictional force $\int ds \Gamma_{2}(t-s)\dot{R}(s) = \eta(R)\dot{R}(t)$ 
where (\eqref{eq:fric_kerR2_2} of \cite{supp}) 
\begin{equation}
	\eta(R)=\pi\lcb\frac{d\Delta(R)}{dR}\frac{\lsb \ene_{F}-\tilde{\ene}_{a}(R)\rsb}{\Delta(R)}\
	+\frac{d\tilde{\ene}_{a}(R)}{dR}\rcb^{2}\rho_{a}^{2}(\ene_{F},R),
	\label{eq:fric}
\end{equation}
is the electronic friction coefficient.
Here, $\rho_{a}(\ene_{F},R)=\frac{1}{\pi}\frac{\Delta(R)}{[\ene_{F}-\tilde{\ene}_{a}(R)]^{2}+\Delta^{2}(R)}$ 
is the particle local density of states.
\eqref{eq:fric} agrees with different forms of $\eta(R)$ that have been derived over the years 
with varying levels of 
generalizations\cite{Nourtier1977,Makoshi1983,Brako1981,Schonhammer1980,HGT1995,Plihal1998,Trail2003,Mizielinski2005,Dou2017}.
\eqref{eq:Langevin_1} then reduces to
\begin{equation}
	\begin{aligned}
	m\ddot{R}(t) + \tilde{V}'(R(t)) \
	+\gamma_{eff}(R)\dot{R}(t) =\xi(t),
\end{aligned}
\label{eq:Langevin_2}
\end{equation}
where $\gamma_{eff}(R)=\gamma+\eta(R)$ and $\xi(t)=\xi_{\gamma}(t)+\xi_{\eta}(t)$
is the effective random force with a corresponding autocorrelation function
$K(t-s)=K_{\gamma}(t-s)+K_{\eta}(t-s)$, where
\begin{equation}
	\begin{aligned}
	K_{\gamma}(t-s) &=  \gamma\int_{0}^{\infty}d\Omega\Omega\
		\coth\left(\frac{\Omega}{2k_{B}T}\right)\cos\Omega(t-s)\\
		K_{\eta}(t-s) &= \eta(R)\int_{0}^{\infty}d\omega\omega\
		\coth\left(\frac{\omega}{2k_{B}T}\right)\cos\omega(t-s).
\end{aligned}
\label{eq:force_cor_2}
\end{equation}

\subsection{Vibrational Relaxation of H on metal surfaces}

\begin{figure}
  \centering
  \mbox{
	  \subfloat[\label{a}]{\includegraphics[scale=0.32]{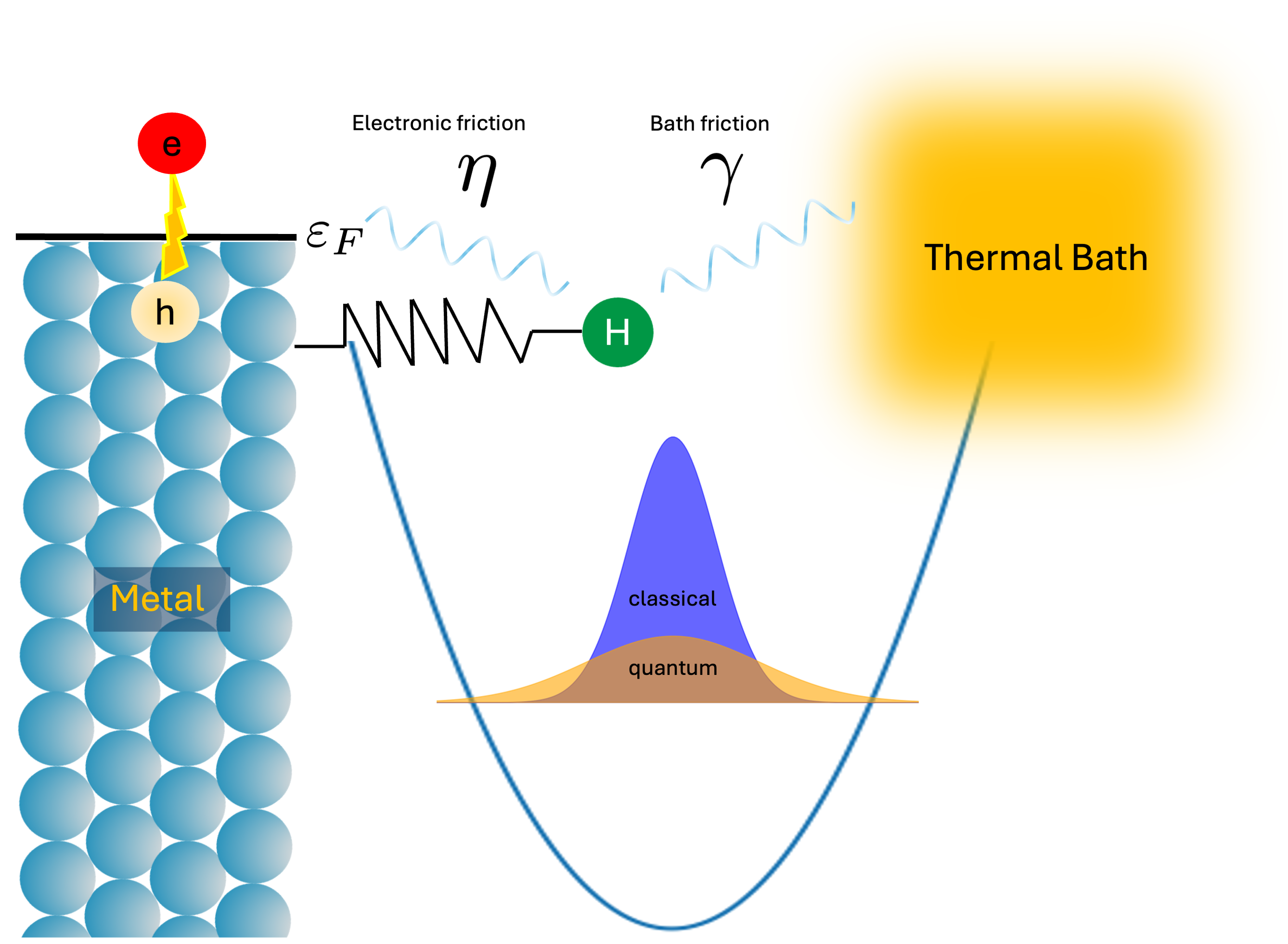}\label{fig:system}}\quad
	  \subfloat[\label{b}]{\includegraphics[scale=0.42]{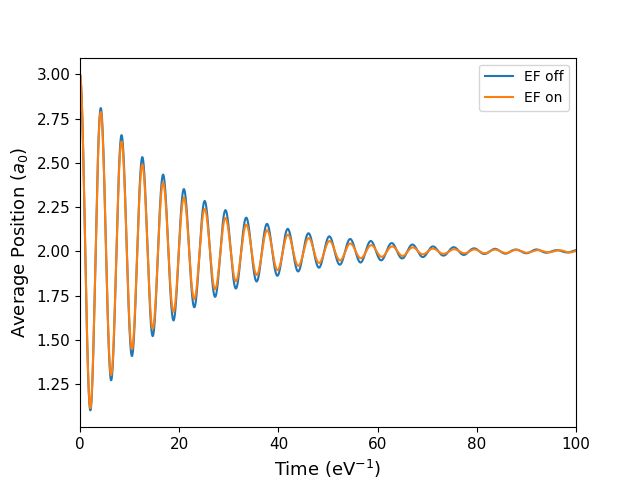}\label{fig:h_av_pos_EF}}\quad
  }
  \mbox{
	  \subfloat[\label{a}]{\includegraphics[scale=0.42]{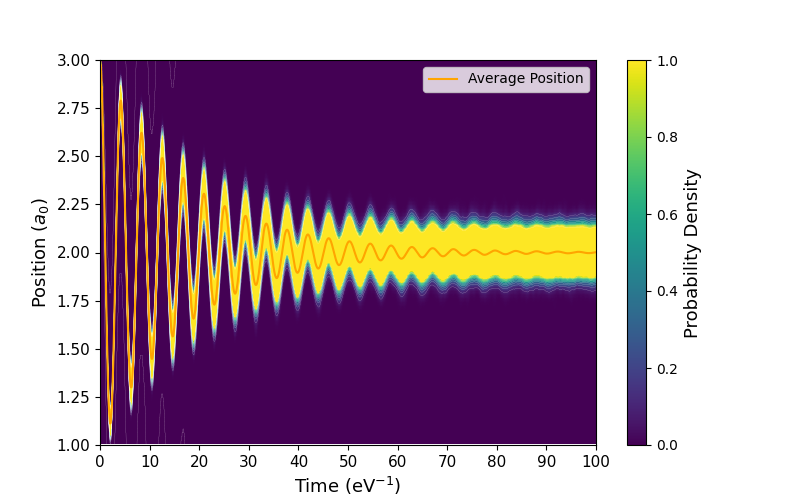}\label{fig:h_pos_c}}\quad
	  \subfloat[\label{b}]{\includegraphics[scale=0.42]{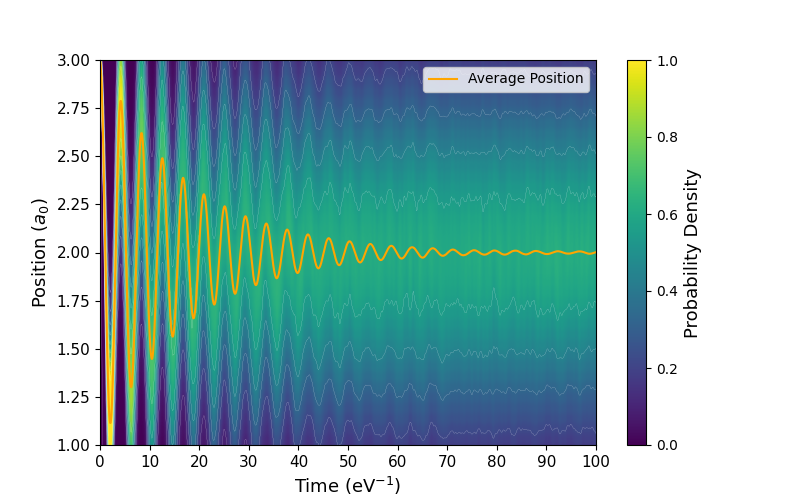}\label{fig:h_pos_q}}\quad
  }
  \captionsetup{justification=raggedright,singlelinecheck=false}
  \caption{(a) A hydrogen adsorbed on a metal surface displaced away from its equilibrium
	  position $R_{0}=2.0~a_{0}$ by $R_{d}=1.0~a_{0}$. 
	  The kinetic energy is dissipated into the electronic and phonon degrees of freedom 
	  via frictions coefficients $\eta$ and $\gamma$, respectively. 
	  (b) Ensemble averaged position of H evolving over time with and without the 
	  electronic friction (EF) effects. (c) Probability distribution of the positions 
	  of H calculated with the classical kernel $K(t-s)=2\gamma_{eff}k_{B}T\delta(t-s)$.
	  (d) The same as in (c) but using the quantum kernel 
	  $K(t-s)=2\gamma_{eff}k_{B}T_{eff}\delta(t-s)$.
  }
  \label{fig:system_pos}
\end{figure}

We first demonstrate the vibrational relaxation of a hydrogen (H) atom 
adsorbed on a metal surface using a harmonic potential of the 
form $V(R)=\frac{1}{2}m\Omega_{0}^{2}(R-R_{0})^{2}$, 
as depicted in \figref{fig:system}. We assume that H is located initially at a 
position away from its equilibrium position $R_{0}=2.0~a_{0}$ ($a_{0}$ is the Bohr radius) 
by $R_{d}=1.0~a_{0}$. H moves towards the equilibrium position in an oscillatory manner, 
while dissipating energy, via $\gamma$ and $\eta(R)$, into the excitation of the 
thermal bath phonons and the creation of coherent e-h pairs in the metal, respectively. 
Because of smallness of the amplitude of H oscillation, $\eta(R)$ has been set 
to be a constant $\eta=\tau^{-1}$. $\tau=76~\mathrm{eV}^{-1}$ ($\approx 50$ fs)
is set within the typical relaxation time of e-h excitations\cite{Alducin2017,Auerbach2021}.
The quantum effects are considered by the quantum fluctuations of $\xi(t)$. 
For a practical reason, we adopt a Gaussian white noise approximation 
to the dissipation kernel K(t-s) as introduced recently 
by Furutani and Salasnich\cite{Furutani2021,Furutani2023}, the 
use of which was validated in the low friction limit for a harmonic potential system. 
This amounts to rewriting $K(t-s)=2\gamma_{eff}k_{B}T_{eff}\delta(t-s)$, 
where $T_{eff}=\frac{\Omega_{0}}{2k_{B}}\coth\left( \frac{\Omega_{0}}{2k_{B}T}\right)$.
We solved \eqref{eq:Langevin_2} ($m=1$) using the symplectic BAOAB method\cite{Leimkuhler2013}
and obtained the averaged quantities $\langle Q \rangle$ 
by $\langle Q \rangle = \frac{1}{N}\sum_{i}^{N}Q_{i}$, 
where $N=50000$ is the number of simulations. 
In the simulations, we set $\gamma=0.1$ eV, $\Omega_{0}=1.5$ eV and $T=100~K$.
\figref{fig:h_av_pos_EF} shows that
the main effect of the electronic friction is to induce slightly 
faster vibrational relaxation by enhancing the damping of the quantum dynamics. 
\figref{fig:h_pos_c} and \figref{fig:h_pos_q}
which are obtained using the quantum kernel $K(t-s)=2\gamma_{eff}k_{B}T_{eff}\delta(t-s)$
and the classical one $K(t-s)=2\gamma_{eff}k_{B}T\delta(t-s)$, respectively, 
show the probability distributions on top of the plot of the average. 
The peaks of density are not normalized and are plotted with $N=5000$ 
between 0 and 1.0 for illustrative purposes. 
The figures show significant discrepancies existing between 
the classical and quantum simulations: although the average positions look similar, 
the probability distribution is much narrower and sharper for the classical case 
than for the quantum case. For the quantum case, the probability distribution 
may be interpreted as $|\Psi(R)|^{2}$ where $\Psi(R)$ is the wavefunction 
of the H nuclear position. In terms of this, one can recognize that the 
quantum fluctuations, as introduced by using $T_{eff}$ instead of $T$, 
affected the spread of wavefunction $|\Psi(R)|^{2}$ as a purely quantum effect.

\subsection{Proton Discharge on metal electrodes}

\begin{figure}
  \centering
  \mbox{
	  \subfloat[\label{}]{\includegraphics[scale=0.32]{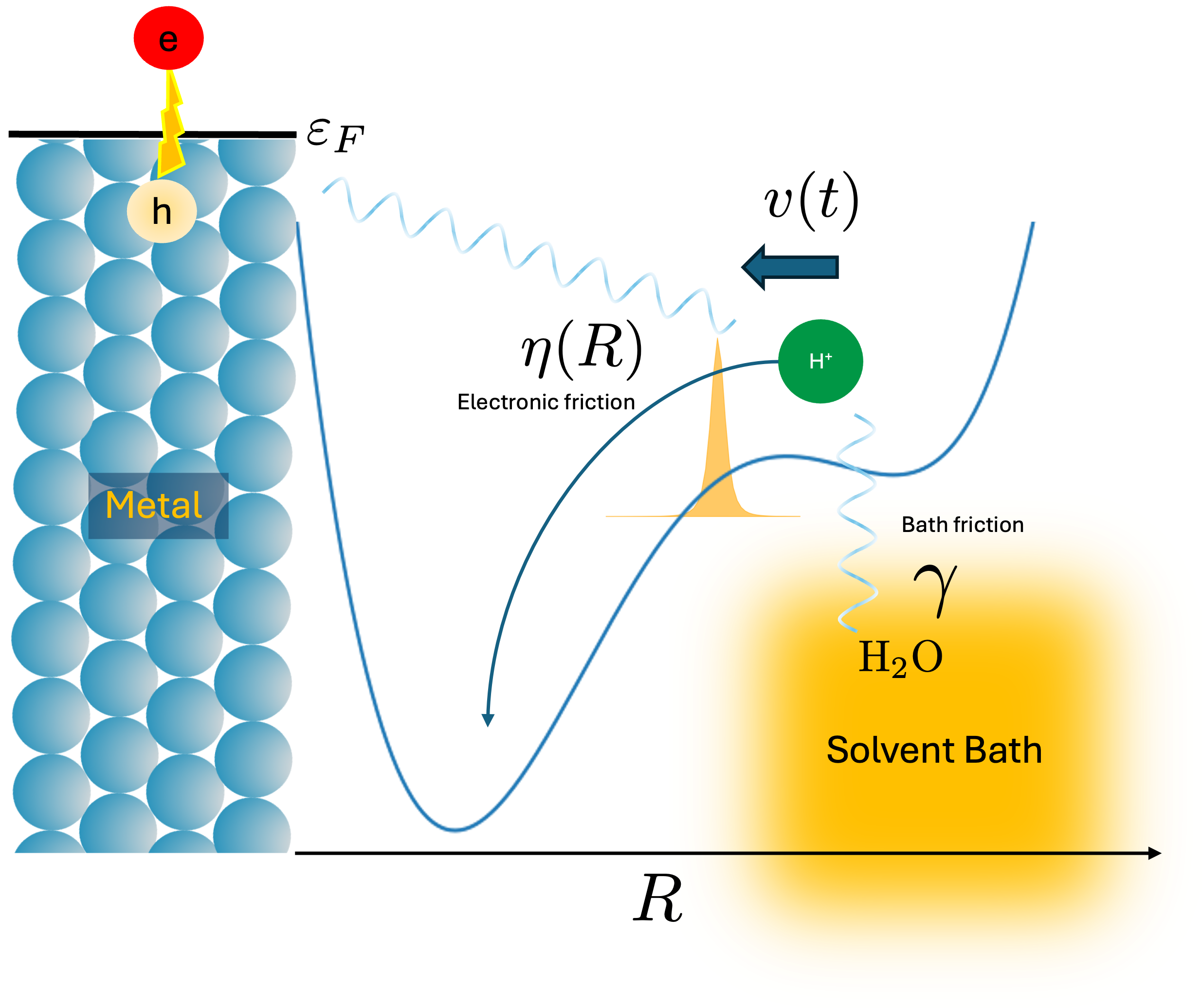}\label{fig:pcet_system}}\quad
	  \subfloat[\label{}]{\includegraphics[scale=0.35]{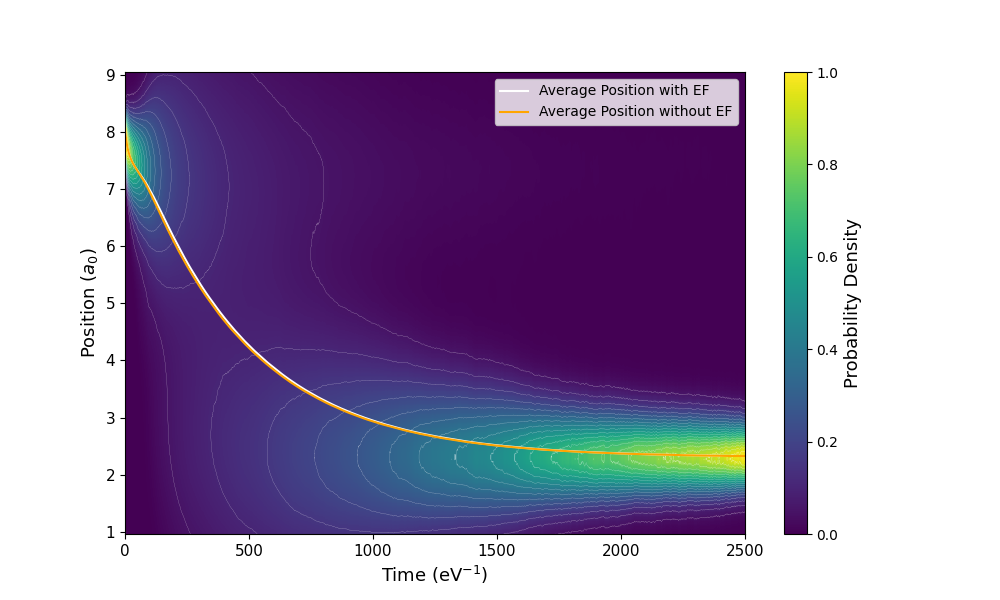}\label{fig:dw_av_pos_EF}}\quad
  }
  \mbox{
	  \subfloat[\label{}]{\includegraphics[scale=0.42]{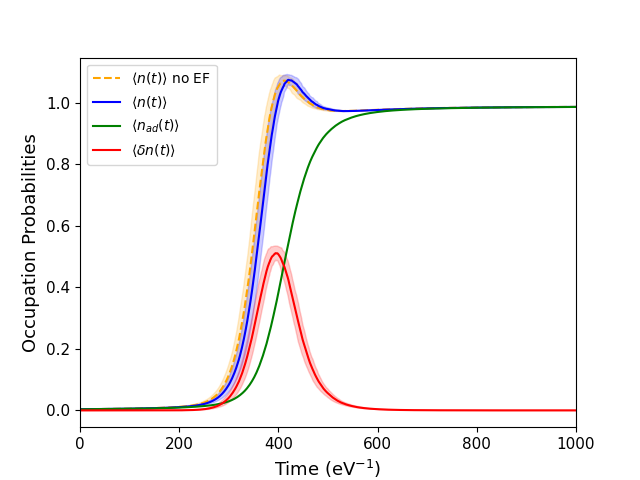}\label{fig:occu_EF}}\quad
	  \subfloat[\label{}]{\includegraphics[scale=0.42]{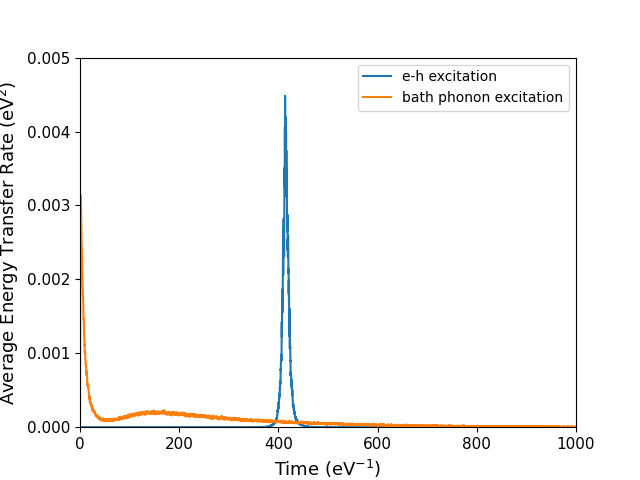}\label{fig:etran_EF}}\quad
  }
  \captionsetup{justification=raggedright,singlelinecheck=false}
  \caption{(a) Schematic representation of the dynamics of $\mathrm{H}^{+}$ discharge on metal electrodes.
	  $\mathrm{H}^{+}$ deposition on metal is described by an excited state PES
	  $V_{H^{+}}(R)=V(R)+\tilde{\ene}_{a}(R)$ where 
	  $V(R)=A\lcb \lsb s(R-\sigma)\rsb^{4} - \lsb s(R-\sigma)\rsb^{2} + a(R-\sigma)\rcb$
	  is a standard double well function. The parameters $A=1.5$ eV, $s=0.25~a_{0}^{-1}$, $\sigma=5.0~a_{0}$ 
	  and $a=-0.05~a_{0}^{-1}$ are chosen such that the assymetric $V_{H^{+}}(R)$ 
	  well depths are at $R_{m1}=8.0~a_{0}$ and $R_{m2}=2.0~a_{0}$.
	  (b) The time evolution of the average position and probability density. 
	  (c) Electron occupation probabilities of $\mathrm{H}^{+}$ as functions of time. The 
	  adiabatic and nonadiabatic components of occupation are also shown along with the effects
	  of electronic friction. (d) The average rate of energy dissipation to 
	  the electron-hole pairs and bath phonons excitations as functions of time.
  }
  \label{fig:system_pcet}
\end{figure}

Let us now consider the dynamics of electrochemical proton discharge which occurs 
on metal electrodes in the presence of solvents.
We study the simplest case corresponding to the Volmer step\cite{Levich1970}
represented as 
$\mathrm{H_{3}O^{+}}+e^{-}\rightarrow \mathrm{H^{*}}+\mathrm{H_{2}O}$, which 
involves dissociation of $\mathrm{H^{+}}$ from $\mathrm{H_{3}O^{+}}$ and 
subsequent deposition on electrode surface as $\mathrm{H^{*}}$ after accepting
an electron. As schematically presented in \figref{fig:pcet_system}
we use a potential energy surface $V_{H^{+}}(R)$ with double minima ($R_{m1}=8.0~a_{0}$ 
and $R_{m2}=2.0~a_{0}$) and a barrier ($R_{B}=6.02~a_{0}$) to simulate 
a movement of $\mathrm{H}^{+}$ across $R_{B}$. The electronic crossing point $R_{c}$ at which
$V_{H^{+}}(R_{c})=V(R_{c})$  or equivalently, $\tilde{\ene}_{a}(R_{c})=\ene_{F}=0$ 
is given by $R_{c}=4.67~a_{0}$ (see Figure 2a caption for details). 
The movement of H experiences a collective drag 
$\gamma_{eff}(R)=\gamma+\eta(R)$ due to excitation of the solvent bath phonons ($\gamma$)
and creation of the e-h pairs in the metali ($\eta(R)$).
To describe the $R$-dependence of $\eta(R)$, we represent $\Delta(R)$ and $\ene_{a}(R)$
as Gaussian functions\cite{Yoshimori1986,Nakanishi1988,supp} whose parameters were 
chosen so that their maxima are located at $R_{m1}$. 
We set $\lambda=\frac{1}{2}G^{e}_{solv}\approx 0.7$ eV, where $G^{e}_{solv}$ is the
electron solvation energy in water\cite{Zhan2003}.
We note that in \eqref{eq:fric}, $\eta(R_{c})$ diverges in proportion 
to $\lsb\tilde{\ene}_{a}^{\prime}(R)/\Delta(R_{c})\rsb^{2}$
for $\Delta(R_{c})\to 0$, but 
for our chosen parameters, $\eta(R)$ sharply peaked at $R_{c}$ (see \figref{fig:pcet_system}).
We carried out $N=500000$ Langevin simulations at $T=300$ K for the duration of 
$t=2500~\mathrm{eV}^{-1}$ ($\approx1600$ fs),
using the stochastic variable $\xi(t)$ that satisfies 
$\langle\xi(t)\xi(s)\rangle=2\gamma_{eff}(R)k_{B}T\delta(t-s)$,
where we set $\gamma=2.5$ eV ($\approx 3.8~\mathrm{fs}^{-1}$).
$\mathrm{H}^{+}$ was placed initially at $R_{m1}$ with a zero initial velocity.
The resulting  $\langle R(t) \rangle$ depicted in \figref{fig:dw_av_pos_EF} shows that
EF slightly delays the crossing towards $R_{m2}$ due to $\eta(R)$ being 
significant only within a narrow range of $R$.
The distribution of the H position initially peaked at 
$R_{m1}$ gradually approaches $R_{m2}$  without destroying the peak structure for long $t$. 
This suggests that recrossing back to $R_{m1}$ side of the PES is highly unlikely,
and can be discerned from the extremely assymetric nature of $V_{H^{+}}(R)$.
Using $\langle R(t) \rangle$ and $\langle v(t) \rangle$, we evaluated the 
the time evolution of the nonadiabatic electron occupation 
$\langle n_{a}(t) \rangle$ as shown in \figref{fig:occu_EF}
for cases with and without EF.
In slow particle motion limit, $\langle n_{a}(t) \rangle$ 
can be decomposed\cite{Arguelles2024} as
$\langle n_{a}(t) \rangle=\langle n_{a}^{ad}(t) \rangle + \langle \delta n_{a}(t) \rangle$,
where, $\langle n_{a}^{ad}(t) \rangle$ 
and $\langle \delta n_{a}(t) \rangle\propto\langle v(t)\rangle$ 
are the adiabatic and nonadiabatic components, respectively.
In both cases, $\langle n_{a}(t) \rangle$ initially increases exponentially 
with time and then shows highly oscillatory overpopulation ($\langle n_{a}(t) \rangle > 1$)
as manifested by the Langevin velocities. The amplitude of the oscillation 
is described by the shaded region, of which the width has been plotted in proportion 
to the variance of the density distribution. The oscillatory overpopulation quickly 
decays to the equilibrium after crossing the Fermi level mainly due to the 
nonadiabatic effect, or by $\langle \delta n_{a}(t) \rangle$. 
Thereafter, the charge transfer occurs over a short time, as represented 
by the partial occupation occurring as early as $t\approx300 eV^{-1}$ (197 fs). 
It is worth pointing out that $\langle n_{a}(t) \rangle$ deviates 
from $\langle n_{a}^{ad}(t) \rangle$ even at slow speeds, whereas the effect 
of the electronic friction (EF), is seen to slightly delay the change in 
the occupation of the energy level. 
The delay effect can be also seen in \figref{fig:dw_av_pos_EF} wherein 
the time evolution of $\langle R(t) \rangle$ is suppressed by an additional drag due to EF.
Finally, we show in \figref{fig:etran_EF} 
the average energy dissipation rate $\dot{\bar{E}}_{\gamma}(t)=\gamma \langle v(t) \rangle^{2} $ 
and $\dot{\bar{E}}_{\eta}(t) = \eta(R) \langle v(t) \rangle^{2}$ to the 
bath phonon and e-h pair excitations, respectively,
as functions of $t$. The dissipation to the bath phonon excitations is seen to take effect
immediately after the particle starts moving and persists over a wide range of $t$. 
$\dot{\bar{E}}_{\gamma}(t)\to 0 $ as $\langle v(t) \rangle^{2}\to 0$ at $t\to\infty$.
In contrast, $\dot{\bar{E}}_{\eta}(t)$ is highly localized within the time interval when
$\tilde{\ene}_{a}(t)$ crosses FL and $\eta(R)$ is peaked. In other words, the
energy dissipation towards e-h excitations only occurs within this short range of $t$.
The total average dissipated energy $\bar{E}_{t}=\bar{E}_{\gamma}+\bar{E}_{\eta}=0.145$ eV 
where the individual contributions are
$\bar{E}_{\gamma} = 0.091$ eV and $\bar{E}_{\eta} = 0.054$ eV. 
$\bar{E}_{t}$ can be thought as the (average) energy gained by the particle from thermal
fluctuations generated by $K(t-s)$, allowing it to overcome the potential barrier of $0.144$ eV
at $R_{B}$ and move to $R_{m2}$.

\section{Discussions and Conclusions}
Using the influence function method\cite{Feynman2000}, we have derived a non-phenomenological 
equation of motion for a particle accounting for simultaneous interactions 
with bosonic and fermionic thermal reservoirs. As an example of the fermionic reservoir, 
we take electrons in a metal electrode, which dissipate the particle energy into the
creation of electron-hole pairs thereby causing electronic friction (EF). 
An explicit expression for the local dissipation kernel (Markovian kernel) 
is given in the limit of slow particle motion providing thereby a way to compute 
the rate of energy transfer via a stochastic simulation. For demonstration purposes, 
we applied the framework to prototypical electrochemical systems where a hydrogen (H) atom 
moves in contact with a metal electrode and a solvent and investigated 
the interplay of the reservoirs using two setups: 
(1) quantum vibrational relaxation of a hydrogen (H) confined on a metal surface and 
(2) solvated electrochemical proton discharge. In the former case we have seen 
that the electronic friction enhances the vibrational friction, and in addition, 
quantum fluctuation effect manifests as a spread of the wave function of H within 
the Gaussian white noise approximation\cite{Furutani2023}. In the latter case, 
we have seen that, contrary to the phononic friction, the electronic friction 
is active at a short time interval of electronic level crossing and contributes 
to delaying of the electron and proton transfers. On the other hand, 
the average energy dissipation is comparable between the two reservoirs
despite the electronic friction being local near the level crossing.
This work marks the first explicit consideration of friction from b
oth solvent modes and electronic excitations in the metal during 
electrochemical proton discharge. Although initially motivated by electrochemistry, 
the presented framework is be broadly applicable to systems 
where interactions with multiple thermal baths play a significant role.

\section{Acknowlegments}
This research is supported by the New Energy and Industrial Technology Development Organization 
(NEDO) project, MEXT as “Program for Promoting Researches on the Supercomputer Fugaku” 
(Fugaku battery \& Fuel Cell Project) 
(Grant No. JPMXP1020200301, Project No.: hp220177, hp210173, hp200131), 
Digital Transformation Initiative for Green Energy Materials (DX-GEM)
and JSPS Grants-in-Aid for Scientific Research (Young Scientists) No. 19K15397. 
Some calculations were done using the supercomputing facilities of the 
Institute for Solid State Physics, The University of Tokyo.
\bibliography{Ref_PI.bib}

\makeatletter\@input{yy.tex}\makeatother
\end{document}


\title[]{Supplementary File: \\
         Dynamics with Simultaneous Interactions with Bosonic and Fermionic Reservoirs}
\author{Elvis F. Arguelles}
\affiliation{Institute for Solid State Physics, The University of Tokyo}
\author{Osamu Sugino}
\affiliation{Institute for Solid State Physics, The University of Tokyo}

\date{\today}

\maketitle
\section{Composite Bosonic Bath Hamiltonian}
\label{sup:tot_Ham}
The interacting composite reservoirs Hamiltonian is expressed as
\begin{equation}
	H_{\mathcal{R}} = \sum_{j}\left(\frac{P_{j}^{2}}{2M_{1}}+\frac{1}{2}M_{1}\Omega^{2}_{1j}X_{j}^{2}\right)\
	+ \sum_{q}\left(\frac{P_{q}^{2}}{2M_{2}}+\frac{1}{2}M_{2}\Omega^{2}_{2q}X_{q}^{2}\right)\
	+\sum_{jq}\beta_{jq}X_{j}X_{q},
	\label{eq:reservoirs_12}
\end{equation}
where, $\beta_{jq}$ is the coupling strength,
$X_{j(q)}$ are the coordinates of harmonic oscillators and  
$P_{j(q)}= -i\frac{\partial}{\partial X_{j(q)}}$ 
are the corresponding momenta.
Introducing a transformation 
$\tilde{X}_{j}=X_{j}+\sum_{q}\frac{\beta_{jq}X_{q}}{M_{1}\Omega_{1j}^{2}}$
results in
\begin{equation}
	H_{\mathcal{R}} = \sum_{j}\left(\frac{\tilde{P}_{j}^{2}}{2M_{1}}+\frac{1}{2}M_{1}\Omega^{2}_{1j}\tilde{X}_{j}^{2}\right)\
	+ \sum_{q}\left(\frac{P_{q}^{2}}{2M_{2}}+\frac{1}{2}M_{2}\tilde{\Omega}^{2}_{2q}X_{q}^{2}\right),
	\label{eq:reservoirs_12_1}
\end{equation}
where
$\tilde{\Omega}_{2q}=\left(\Omega_{2q}^{2}-\sum_{j}\frac{\beta_{jq}^{2}}{M_{1}M_{2}\Omega_{1j}^{2}}\right)^{1/2}$. In second quantized notation we arrive at
\begin{equation}
H_{\mathcal{R}} = \sum_{j}\Omega_{1j}B_{j}^{\dagger}B_{j}
	+ \sum_{q}\Omega_{2q}\bc\ba,
	\label{eq:reservoirs_12_2}
\end{equation}
where the dressing in $B_{j}^{\dagger}(B_{j})$ operators due to the inter-reservoir
coupling is implied. Further we drop the tilde on the frequency in the main text.

\section{Thermal Average of the Influence Functional}
\label{sup:bos_trace}
Let us now evaluate the thermal average in \eqref{eq:F12_xy}. 
For brevity we let $Q\equiv\mathscr{F}_{\alpha}[x,y]$
\begin{equation}
	\begin{aligned}
		Q &=\Bigg< \tilde{\mathcal{T}}\exp\lcb i\int dt\sum_{j}\lsb c_{j}(x(t))e^{-i\Omega_{1j}t}B_{j}+c_{j}^{*}(x(t))e^{i\Omega_{1j}t}B_{j}^{\dagger}\rsb\rcb\\
&\times\mathcal{T}\exp\lcb-i\int ds
\sum_{j}\lsb h_{j}(y(s))e^{-i\Omega_{1j}t}B_{j}+h_{j}^{*}(y(s))e^{i\Omega_{1j}t}B_{j}^{\dagger}\rsb\rcb\Bigg>,
\end{aligned}
	\label{eq:Q_xy}
\end{equation}
Let us first drop the summation over $j$ and "1" in $\Omega_{1}$ and
express the anti-time ordered expression $\tilde{\mathcal{T}}(\ldots)$ as\cite{Newns1985}
$(\hbar = 1)$
\begin{equation}
	\exp\left(i\int_{-\infty}^{\tau}dt c_{t}^{*}e^{i\Omega t}B^{\dagger}\right) \
		\exp\left(i\int_{-\infty}^{\infty}d\tau c_{\tau}e^{-i\Omega \tau}B\right) \
		\exp\left(i\int_{\tau}^{\infty}dt c_{t}^{*}e^{i\Omega t}B^{\dagger}\right).
	\label{eq:anti_ord}
\end{equation}
To reorder the last two factors, we make use of $e^{A}e^{B}=e^{B}e^{A}e^{[A,B]}$ with
\begin{equation}
	[A,B] = -\int_{-\infty}^{\infty}d\tau c_{\tau}e^{-i\Omega \tau}\
	\int_{\tau}^{\infty}dt c_{t}^{*}e^{i\Omega t}\equiv -P.
	\label{eq:commute1}
\end{equation}
Expression \ref{eq:anti_ord} becomes
\begin{equation}
		\exp\left(i\int_{-\infty}^{\infty}dt F_{t}^{*}e^{i\omega t}b_{\alpha}^{\dagger}\right) \
		\exp\left(i\int_{-\infty}^{\infty}d\tau F_{\tau}e^{-i\omega \tau}b_{\alpha}\right)\exp(-P).
	\label{eq:anti_ord2}
\end{equation}
Similarly for the time-ordered expression $\mathcal{T}(\ldots)$ (earliest time to the right),
\begin{equation}
		\exp\left(-i\int_{\tau}^{\infty}dt h_{t}^{*}e^{i\Omega t}B^{\dagger}\right)
		\exp\left(-i\int_{-\infty}^{\infty}d\tau h_{\tau}e^{-i\Omega \tau}B\right) \
		\exp\left(-i\int_{-\infty}^{\tau}dt h_{t}^{*}e^{i\Omega t}B^{\dagger}\right) \
	\label{eq:ti_ord}
\end{equation}
with
\begin{equation}
	[A,B] = -\int_{-\infty}^{\infty}d\tau h_{\tau}e^{-i\Omega \tau}\
	\int_{-\infty}^{\tau}dt h_{t}^{*}e^{i\Omega t} \equiv -D,
	\label{eq:commute3}
\end{equation}
so that
\begin{equation}
		\exp\left(i\int_{-\infty}^{\infty}dt h_{t}^{*}e^{i\Omega t}B^{\dagger}\right) \
		\exp\left(i\int_{-\infty}^{\infty}d\tau h_{\tau}e^{-i\Omega \tau}B\right)\exp(-D).
	\label{eq:ti_ord2}
\end{equation}
For compact notation, let us define
\begin{equation*}
	\begin{aligned}
	\int_{-\infty}^{\infty}d\tau c_{\tau}e^{-i\Omega \tau} \equiv c_{\Omega} \\
	\int_{-\infty}^{\infty}d\tau h_{\tau}e^{-i\Omega \tau} \equiv h_{\Omega}.
\end{aligned}
	\label{eq:redefined}
\end{equation*}
The thermal average is then written as
\begin{equation}
	Q = \Big< e^{-(P+D)}e^{\left(ic^{*}_{\Omega}B^{\dagger}+ic_{\Omega}B\right)} \
	e^{\left(-ih^{*}_{\Omega}b^{\dagger}+ih_{\Omega}B\right)}\Big>,
	\label{eq:thermal_ave2}
\end{equation}
which can also be written in terms of trace over the boson states
\begin{equation}
	\begin{aligned}
		Q &= e^{\beta\Omega_{bos}}\mathrm{Tr}\left[e^{-\beta\sum_{j}\Omega_{j}n_{j}} \
e^{-(P+D)}e^{\left(ic^{*}_{\Omega}B^{\dagger}+ic_{\Omega}B\right)} \
	e^{\left(-ih^{*}_{\Omega}B^{\dagger}+ih_{\Omega}B\right)}\right], \\
		 &= \Pi_{j}e^{\beta\Omega_{j}}\sum_{n_{j}=0}^{\infty}e^{-\beta\sum_{j}\Omega_{j}n_{j}}\
		 \langle n_{j} | e^{-(P_{j}+D_{j})}e^{\left(ic^{*}_{\Omega_{j}}B_{j}^{\dagger}+ic_{\Omega_{j}}B_{j}\right)} \
		 e^{\left(-ih^{*}_{\Omega_{j}}B_{j}^{\dagger}+ih_{\Omega_{j}}B_{j}\right)}| n_{j} \rangle \\
	        &=\Pi_{j}\mathcal{Q}_{j},
\end{aligned}
	\label{eq:thermal_ave3}
\end{equation}
where we have reintroduced (temporarily) the $j$ dependence for clarity with the prefactor
\begin{equation}
	e^{\beta\Omega_{j}} = \left(\sum_{n_{j}=0}^{\infty} e^{-\beta n_{j}\Omega_{j}} \right)^{-1} = 1-e^{-\beta\Omega_{j}}.
	\label{eq:prefact}
\end{equation}
Let us again drop $j$s in such a way that
\begin{equation}
	\mathcal{Q} =e^{-(P+D)}\left( 1-e^{-\beta\Omega}\right)\sum_{n=0}^{\infty}e^{-\beta\Omega n}\,
	\langle n |e^{ic^{*}_{\Omega}B^{\dagger}}e^{ic_{\Omega}B} \
	e^{-ih^{*}_{\Omega}B^{\dagger}}e^{-ih_{\Omega}B}| n \rangle
	\label{eq:thermal_ave4}
\end{equation}
The next step is to rearrange the operators so that the annihilation operators are on the right
and the creation operators are on the left. This means we need to reorder the two operators in the 
middle of \eqref{eq:thermal_ave4}. 
Since they do not commute, the rearrangement yields a phase factor giving
\begin{equation}
	e^{ic_{\Omega}B} e^{-ic^{*}_{\Omega}B^{\dagger}}=\
	e^{-ih^{*}_{\Omega}B^{\dagger}}\left[e^{ih^{*}_{\Omega}B^{\dagger}}e^{ic_{\Omega}B}e^{-ih^{*}_{\Omega}B^{\dagger}}\right].
	\label{eq:rear_ops}
\end{equation}
The factors in the bracket may be written as
\begin{equation}
	e^{ih^{*}_{\Omega}B^{\dagger}}e^{ic_{\Omega}B}e^{-ih^{*}_{\Omega}B^{\dagger}}=\
	e^{ic_{\Omega}B}\exp\left(h^{*}_{\Omega}c_{\Omega}\right),
	\label{eq:rear_ops1}
\end{equation}
such that
\begin{equation}
	\begin{aligned}
	\mathcal{Q}=e^{-\left(P+D-h^{*}_{\Omega}c_{\Omega}\right)}\
	\left( 1-e^{-\beta\Omega}\right)\sum_{n=0}^{\infty}e^{-\beta\Omega n}\
	\langle n |e^{\left(ic^{*}_{\Omega}-ih^{*}_{\Omega}\right)B^{\dagger}}\
	e^{\left(ic_{\Omega}-ih_{\Omega}\right)B}| n \rangle.
\end{aligned}
	\label{eq:thermal_ave5}
\end{equation}
Let $u=-\left(ic^{*}_{\Omega}-ih^{*}_{\Omega}\right)$ and $v=ic_{\Omega}-ig_{\Omega}$ and write
\eqref{eq:thermal_ave5} as
\begin{equation}
	\mathcal{Q}=\
	e^{-\left(P+D-h^{*}_{\Omega}c_{\Omega}\right)}\
	\left( 1-e^{-\beta\Omega}\right)\
	\sum_{n=0}^{\infty}e^{-\beta\Omega n}\langle n |e^{vB^{\dagger}}e^{-u B} | n \rangle,
	\label{eq:thermal_ave6}
\end{equation}
and expand $e^{-u B}$ in power series so that
\begin{equation}
	e^{-u B} | n \rangle = \sum_{l=0}^{\infty}\frac{(-u)^{l}}{l!}B^{l}| n \rangle.
	\label{eq:anni_power}
\end{equation}
For annihilation operator $B$,
\begin{equation}
	B^{l}| n \rangle=\left[\frac{n!}{(n-l)!}\right]^{1/2}| n -l \rangle,
	\label{eq:anni_prop}
\end{equation}
with $B^{l}| n \rangle=0$ for $l>n$. We can terminate the series at $l=n$ and
write \eqref{eq:anni_power} as
\begin{equation}
	e^{-u B} | n \rangle = \sum_{l=0}^{n}\frac{(-u)^{l}}{l!}\left[\frac{n!}{(n-l)!}\right]^{1/2}| n -l \rangle.
	\label{eq:anni_power1}
\end{equation}
Similarly for operator $B^{\dagger}$,
\begin{equation}
	\langle n |e^{vB^{\dagger}} = \sum_{m=0}^{n}\frac{(v)^{m}}{m!}\left[\frac{n!}{(n-m)!}\right]^{1/2} \langle n -m |,
	\label{eq:crea_power1}
\end{equation}
so that 
\begin{equation}
	\begin{aligned}
	\langle n |e^{vB^{\dagger}}e^{-u B}| n \rangle	&= \sum_{l=0}^{n}\sum_{m=0}^{n}\
	\frac{(-u)^{l}}{l!}\frac{(v)^{m}}{m!}\left[\frac{n!}{(n-l)!}\right]^{1/2} \
	\left[\frac{n!}{(n-m)!}\right]^{1/2} \delta_{m=l}\\
	&= \sum_{l=0}^{n}\frac{(-uv)^{l}}{(l!)^{2}}\frac{n!}{(n-l)!} = L_{n}(x).
	\end{aligned}
	\label{eq:expect_n}
\end{equation}
Here, $L_{n}(x)$ is the Laguerre polynomial of order $n$ whose generating function is
\begin{equation}
	\begin{aligned}
	\sum_{n}^{\infty}L_{n}(x)t^{n}=\frac{1}{1-t}e^{-tx/(t-1)}\\
	(1-t)\sum_{n}^{\infty}L_{n}(x)t^{n}=e^{-tx/(t-1)}.
        \end{aligned}
	\label{eq:gen_func}
\end{equation}
By inspection of \eqref{eq:thermal_ave6}, we identify
$t=e^{-\beta\Omega}$ and $\frac{e^{-\beta\Omega}}{e^{-\beta\Omega}-1}=-\frac{1}{e^{\beta\Omega}-1}=-N$,
with
\begin{equation}
	x = -uv = -c_{\Omega}^{*}c_{\Omega} + c_{\Omega}^{*}h_{\Omega}+h_{\Omega}^{*}c_{\Omega}\
	-h_{\Omega}^{*}h_{\Omega}.
	\label{eq:uv}
\end{equation}
We can therefore write
\begin{equation}
	\left( 1-e^{-\beta\Omega}\right)\
	\sum_{n=0}^{\infty}e^{-\beta\Omega n}\langle n |e^{vB^{\dagger}}e^{-u B} | n \rangle =\
	e^{-\left(c_{\Omega}^{*}c_{\Omega} - c_{\Omega}^{*}h_{\Omega}-h_{\Omega}^{*}c_{\Omega}\
	+h_{\Omega}^{*}h_{\Omega}\right)N},
	\label{eq:ther_av_part}
\end{equation}
so that \eqref{eq:thermal_ave6} reads
\begin{equation}
	\mathcal{Q}=\
	e^{-\left(P+D-g^{*}_{\Omega}c_{\Omega}\right)}\
	e^{-\left(c_{\Omega}^{*}c_{\Omega} - c_{\Omega}^{*}h_{\Omega}-h_{\Omega}^{*}c_{\Omega}\
	+h_{\Omega}^{*}h_{\Omega}\right)N}
	\label{eq:thermal_ave7}
\end{equation}
Putting back the $j$'s and rewriting the above in terms of time integrations,
\begin{equation}
	Q=\Pi_{j}\mathcal{Q}_{j}=\exp\left(-\sum_{j}\phi_{j} \right),
	\label{eq:thermal_ave8}
\end{equation}
where
\begin{equation}
	\begin{aligned}
		\phi_{j} &=\int dt\int ds\left\{ e^{-i\Omega_{j}(t-s)}\left[c_{j}[x(t)]c_{j}^{*}[x(s)]\theta(s-t) \
	+ c_{j}^{*}[y(s)]c_{j}[y(t)]\theta(t-s) \right.\right. \\
	&-\left.\left. c_{j}^{*}[y(s)]c_{j}[x(t)]\right](N_{j}+1) \
	- e^{-i\Omega_{j}(s-t)}c_{j}[y(s)]c_{j}^{*}[x(t)]N_{j}\right\}.
	\end{aligned}
	\label{eq:phi}
\end{equation}
Consequently, the influence phase is given by
\begin{equation}
	\Phi[x,y]=-i\ln(Q).
	\label{eq:inf_phase}
\end{equation}

\section{Functional Taylor Expansions}
\label{supp:fte}

\subsection{Particle Action}
\label{supp:S_fte}
We first perform the functional Taylor expansion
for the particle action $S_{p}[R,r]=-[S_{p}(R+r/2)-S_{p}(R-r/2)]$.
Expanding up to $\mathcal{O}(\tr^{3})$ 
\begin{equation}
	S_{p}(R+r/2)=S_{p}(R,0)+\frac{1}{2}\int dt_{1}\frac{\delta S_{p}[r]}{\delta r(t_{1})}\
	\Big|_{r=0}r(t_{1})+\frac{1}{4}\frac{1}{2!}\int dt_{1}\int dt_{2}\
	\frac{\delta^{2}S_{p}[r]}{\delta r(t_{1})\delta r(t_{2})}\Big|_{r=0}\
	r(t_{1})r(t_{2})+\mathcal{O}(r^{3}).
	\label{eq:S_fte}
\end{equation}
Similar expansion can be performed for 
$S_{p}(R-r/2)$. These result in
\begin{equation}
	\begin{aligned}
		S_{p}(R+r/2) &=\int dt \left[\frac{1}{2}m\dot{R}^{2}+\frac{1}{2}m\dot{R}\dot{r} \
	+\frac{1}{8}m\dot{r}^{2} \right. \\
	&\left. - V(R) -\frac{1}{2}V'(R)r -\frac{1}{8}V''(R)r^{2}-\frac{1}{48}V'''(R)r^{3}\right]\\
		S_{p}(R-r/2) &=\int dt \left[\frac{1}{2}m\dot{R}^{2}-\frac{1}{2}m\dot{R}\dot{r} \
	+\frac{1}{8}m\dot{r}^{2} \right. \\
	&\left. - V(R) +\frac{1}{2}V'(R)r -\frac{1}{8}V''(R)r^{2}+\frac{1}{48}V'''(R)r^{3}\right].
\end{aligned}
	\label{eq:sup_S_pxyrexpan}
\end{equation}
Several terms in \eqref{eq:sup_S_pxyrexpan} cancel each other giving
\begin{equation}
		S_{p}(R,r) =-\int dt \left[m\dot{R}\dot{r} \
		-V'(R)r -\frac{1}{24}V'''(R)r^{3}\right], \\
	\label{eq:sup_S_prex}
\end{equation}
which is \eqref{eq:S_prex} of the main text.

\subsection{Influence Phase}
\label{supp:ip_fte}

The influence phase is written as
$\Phi[R,r]=\Pi[R,r]+i\Sigma[R,r]$
where $\Pi\equiv\mathrm{Re}[\Phi]$ and $\Sigma\equiv\mathrm{Im}[\Phi]$ 
are the real and imaginary parts, respectively.
Due to their symmetries, some terms of the expansion vanish. Writing only
the surviving terms after $\mathcal{O}(r^{2})$ expansion gives
\begin{equation}
	\begin{aligned}
	\Pi[R,r] &\approx\frac{1}{2}\int dt_{1}\frac{\delta \Pi[R,r]}{\delta r(t_{1})}\
	\Big|_{r=0}r(t_{1})\equiv \Pi^{(1)}[R,r]\\
	\Sigma[R,r] &\approx \Sigma[R,0]+\frac{1}{4}\frac{1}{2!}\int dt_{1}\int dt_{2}\
	\frac{\delta^{2}S_{p}[r]}{\delta r(t_{1})\delta r(t_{2})}\Big|_{r=0}\
	r(t_{1})r(t_{2})\equiv\Sigma^{(0)}[R,0]+\Sigma^{(2)}[R,r].
\end{aligned}
	\label{eq:phase_com_fte}
\end{equation}
$\Sigma^{(0)}[R,0]$ is irrelevant in the transition amplitude and can be
reintroduced in the full evaluation of $\rho_{red}(t_{f})$.
The remaining expansion terms read 
\begin{equation}
	\Pi^{(1)}(R,r) =-\frac{1}{2}\sum_{j}\int dt \int ds r(t) c'_{j}(R(t))c_{j}^{*}(R(s))[\mathrm{sgn}(t-s)+1]\sin\Omega_{j}(t-s)
	\label{eq:Pi1}
\end{equation}
\begin{equation}
	\Sigma^{(2)}(R,r) =-\frac{1}{4}\sum_{j}\int dt \int ds r(t)r(s)c'^{*}_{j}(R(t))c'_{j}(R(s))\
	\mathrm{coth}\lb\frac{\Omega_{j}}{2k_{B}T}\rb\cos\Omega_{j}(t-s).
	\label{eq:Sigma2}
\end{equation}
To obtain the friction kernel \eqref{eq:fric_ker} in the main text,
we integrate by parts \eqref{eq:Pi1} giving
\begin{equation}
	\begin{aligned}
	\Pi^{(1)}(R,r) &=-\int dt\sum_{j}\frac{c^{\prime}_{j}(R(t))c^{*}_{j}(R(t))}{\Omega_{j}}\\
	&+\int dt \int ds \sum_{j} \frac{c^{\prime}_{j}(R(t))c^{\prime *}_{j}(R(s))}{2\Omega_{j}}\
	[\mathrm{sgn}(t-s)+1]\cos\Omega_{j}(t-s)\dot{R}(s)\\
	&= -\int dt \frac{d \Delta\Omega(R)}{dR} + \int dt \int ds \Gamma(t-s)\dot{R}(s),
\end{aligned}
	\label{eq:Pi1_ibp}
\end{equation}
where $\Delta\Omega(R)\equiv\sum_{j}\frac{|c_{j}(R)|^{2}}{2\Omega_{j}}$ is the potential
shift that renormalizes $V(R)$ and $\Gamma(t-s)$ is the friction kernel (\eqref{eq:fric_ker}).
On the other hand, the integrand in \eqref{eq:Sigma2} gives the autocorrelation function
$K(t-s)$ (\eqref{eq:force_cor}) in the main text.

\section{Stochastic average over the random force}
\label{supp:stoav}
For any propagator $A(t,t')$, the stochastic average of an operator $O$ 
with respect to the Gaussian random force $\xi$ may be written\cite{Furutani2023} as
\begin{equation}
	\langle O \rangle =\frac{\int\Dm[\xi(t)] O[\xi(t)]e^{-\frac{1}{2}}\int dt\int dt'\xi(t)A^{-1}(t-t')\xi(t')}{\int\Dm[\xi(t)]\
	e^{-\frac{1}{2}}\int dt\int dt'\xi(t)A^{-1}(t-t')\xi(t')}.
	\label{eq:O_ave}
\end{equation}
The above allows us to write 
\eqref{eq:J2} in terms of the random force $\xi(t)$ as
\begin{equation}
	\mathcal{J}=\int\Dm R\int\Dm r\Bigg< \exp\lcb i\int dt\lsb m\ddot{R}(t) + \tilde{V}'(R(t)) \
+\int ds\sum_{\alpha}\Gamma_{\alpha}(t-s)\dot{R}(s)-\xi(t)\rsb r(t) \rcb\Bigg>.
	\label{eq:supp_J}
\end{equation}
where $\langle\ldots\rangle$ denotes the stochastic average over $\xi(t)$.
To this end, we first notice that the exponential
term of \eqref{eq:J2} which becomes proportional to 
$\frac{1}{2}\int dt\int ds r(t) K(t-s) r(s) +\frac{1}{2}\int dt\int ds\xi(t)K^{-1}(t-s)\xi(s)$,
where $K^{-1}(t-s)$ is the inverse kernel satisfying
\begin{equation}
	\int dt' K(t-t')K^{-1}(t'-t'')=\delta(t-t'').
	\label{eq:inv_kernel}
\end{equation}
The integration variable may then be shifted as 
$\xi(t) \to \xi(t) + i\int ds K(t-s) r(t)$.
After some simplifications and with the aid of \eqref{eq:inv_kernel}, 
$\frac{1}{2}\int dt\int ds r(t) K(t-s) r(s) +\frac{1}{2}\int dt\int ds\xi(t)K^{-1}(t-s)\xi(s)\
\rightarrow \frac{1}{2}\int dt\int ds\xi(t)K^{-1}(t-s)\xi(s) -i\int dt \xi(t)r(t)$ which
eventually leads to \eqref{eq:supp_J}.
The $r$ integration can then be performed to yield
\begin{equation}
	\mathcal{J}=\int\Dm R \Bigg<\delta\lsb m\ddot{R}(t) + \tilde{V}'(R(t)) \
	+\int ds\sum_{\alpha}\Gamma_{\alpha}(t-s)\dot{R}(s) -\xi(t) \rsb \Bigg>,
\label{eq:J3}
\end{equation}
where the Dirac delta functional gives \eqref{eq:Langevin_1} in the main text.

\section{Perturbed Metal Electron System}

To derive the effective Hamiltonian of the metal electrons perturbed by an approaching particle,
we employ the time-dependent Newns-Anderson-Schmickler (NAS) 
model\cite{Schmickler1986,Arguelles2024} given by
\begin{equation}
	\begin{aligned}
		H_{S}(t) &= \sum_{k}\ene_{k}\cc\ca + \ene_{a}(t)n_{a} + \sum_{k}\lsb V_{ak}(t)\ac\ca + H.c. \rsb \\
	&+\sum_{j}\frac{P_{j}^{2}}{2M} + \sum_{j}\frac{1}{2}M\Omega_{j}^{2}R_{j}^{2} +(Z-n_{a})\sum_{j}\lambda_{j}\Omegj R_{j},
\end{aligned}
	\label{eq:HS}
\end{equation}
where the first term corresponds to the $unperturbed$ Hamiltonian of metal electrons,
$\ene_{a}(t)$ is the particle's time-dependent energy level
and $n_{a}=\ac a$ is its the electron occupancy in terms of $\ac(a)$ creation (annihilation)
operators belonging to the valence electron orbital $|a\rangle$. 
$V_{ak}(t)$ is the transfer integral
between the metal and particle electrons, whose time-dependence we assume
to be separable through $V_{ak}(t)=V_{ak}u(t)$ where $u(t)$ is some time-dependent function
to be specified later. For practical purposes, we assume that the band of the metal electrons
is infinitely wide so that the resonance width
\begin{equation}
	\Delta = 2\pi\sum_{k}|V_{ak}|^{2}\delta(\ene-\ene_{k})
	\label{eq:resowidth}
\end{equation}
is energy independent. 
The last three terms in \eqref{eq:HS} are 
the Hamiltonians of the reservoir represented by a bath of harmonic oscillators,
and the electron-bath oscillators ($phonon$) coupling of strength $\lambda_{j}$.
$Z$ is the charge of the ionized particle ($Z=1$ and $Z=0$ for proton and hydrogen, respectively).
The NAS model describes the electronic interaction between adsorbates and the metal electrode
submerged in a solvent which is represented by a bath of harmonic oscillators. When $Z=0$,
this model reduces to the Anderson-Holstein model where the reservoir is not necessarily
restricted to a solvent.
It is advantageous to eliminate the electron-phonon (e-ph) coupling in
\eqref{eq:HS} using the canonical Lang-Firsov transformation. This is 
conveniently achieved in the second quantized notation of the 
bath phonon momenta and coordinates, i.e., in terms of phonon operators $\Bjc$ and $\Bja$. 
The transformation $\tilde{H}_{S}(t) = e^{S}H_{S}e^{-S}$
where $S=\sum_{j}\lambda_{j}n_{a}\lb \Bjc-\Bja \rb$ results in
\begin{equation}
	\begin{aligned}
		\tilde{H}_{S}(t) &= \tilde{\ene}_{a}(t)n_{a} + \sum_{k}\lsb \tilde{V}_{ak}(t)\ac\ca + H.c. \rsb \\
		   &+ \sum_{j}\Omegj\lsb \Bjc\Bja +Z\lambda_{j}\lb\Bja+\Bjc\rb\rsb\
		   +\sum_{k}\ene_{k}\cc\ca,
	   \end{aligned}
	\label{eq:HSLF}
\end{equation}
where $\tilde{\ene}_{a}(t)=\ene_{a}(t)+(2Z-1)\lambda$ is the particle energy level 
renormalized by $total$ e-ph coupling $\lambda \equiv \sum_{j}\lambda^{2}_{j}\Omega_{j}$ and
$\tilde{V}_{ak}(t)= V_{ak}(t)X^{\dagger}$ is the dressed 
electronic overlap integral with the phonon operator 
$X^{\dagger}=e^{\sum_{j}\lambda_{j}\lb\Bjc-\Bja\rb}$. $\lambda$ is 
called the solvent reorganizational energy in electrochemical literatures.
We replace
$X^{\dagger}\approx\langle X^{\dagger} \rangle =\exp\lsb \sum_{j}\lambda^{2}_{j}\lb N_{j} + 1/2 \rb \rsb$
where $N_{j} = \lb e^{\beta\Omega_{j}}-1 \rb^{-1}$ is the phonon population 
and $\beta = \frac{1}{k_{B}T}$. This is justified as long as 
$V_{ak}\ll \lambda$\cite{Hewson1974,Citrin1977}.
Most physical systems of interest
satisfy this condition.  
As shown in \eqref{eq:HSLF}, the canonical transformation 
leaves the metal electron Hamiltonian unchanged while $H_{R}$
acquires an additional term due to e-ph interaction. 

\subsection{Effective electron-hole excitation Hamiltonian}

Having obtained \eqref{eq:HSLF}, we are now in a position to
derive the effective Hamiltonian of the metal electrons in the presence
of a slowly approaching particle. We follow the procedure outlined by
Brako and Newns\cite{Brako1981,Brako1989}. 
Let us first write the Hamiltonian of metal electrons as
\begin{equation}
	H_{M} = \sum_{k}\ene_{k}\cc(t)\ca(t),
	\label{eq:Hm0}
\end{equation}
The Heisenberg equations of motion for the particle and metal electron operators
are
\begin{equation}
	\dot{a}(t)=i[\tilde{H}_{s},a] = i\tilde{\ene}_{a}(t)a(t) + i\sum_{k}\tilde{V}_{ak}(t)c_{k}(t),
	\label{eq:eom_a}
\end{equation}
\begin{equation}
	\dot{c}_{k}(t)=i[\tilde{H}_{s},c_{k}] = i\ene_{k}(t)c_{k}(t) + i\tilde{V}_{ak}^{*}(t)a(t).
	\label{eq:eom_c}
\end{equation}
The integration of \eqref{eq:eom_c} gives $c_{k}(t)$ which when 
inserted into \eqref{eq:eom_a}
and integrated yields $a(t)$. Insertion of the solution $a(t)$ into
\eqref{eq:eom_c} followed by an integration gives 
\begin{equation}
	\begin{aligned}
	e^{i\ene_{k}t}c_{k}(t)&=e^{i\ene_{k}t}c_{k}(t_{0}) -i\sum_{k'}\int_{t_{0}}^{t}dt' \
	e^{i\ene_{k}t'}\tilde{V}^{*}_{ak}(t')\int_{t_{0}}^{t'}dt'' e^{-i\ene_{k'}t''}\tilde{V}_{ak'}(t'')\\
	&\times \exp\lcb-i\int_{t''}^{t'}d\tau \lsb \tilde{\ene}_{a}(\tau) -i\Delta(\tau)\rsb \rcb e^{i\ene_{k}t}c_{k}(t_{0})\\
	&-i\int_{t_{0}}^{t}dt'e^{i\ene_{k}t'}\tilde{V}^{*}_{ak}(t') \
	\exp\lcb-i\int_{t_{0}}^{t'}dt'' \lsb \tilde{\ene}_{a}(t'') -i\Delta(t'')\rsb \rcb a(t_{0})
\end{aligned}
	\label{eq:eom_c1}
\end{equation}
where $\Delta(t) = \Delta u^{2}(t)$.
To this end the wide band approximation is employed so that \eqref{eq:resowidth} may be written as
$\Delta = 2\pi\sum_{k}|V_{ak}|^{2}\langle X \rangle\langle X^{\dagger} \rangle \delta(\ene-\ene_{k})
\approx 2\pi|V_{ak}|^{2}\rho(\ene_{k})$, where 
$\rho(\ene_{k})$ is the metal electrons density of states (DOS). 
\eqref{eq:eom_c1} is greatly simplified in the limit of slow particle motion. 
In this limit,
\[
	\exp\lcb-i\int_{t''}^{t'}d\tau \lsb \tilde{\ene}_{a}(\tau) -i\Delta(\tau)\rsb \rcb
\]
is nontrivial only if $t'-t'' \lesssim \Delta(t)^{-1}$ so we may approximate it as
\begin{equation}
\exp\lcb-i\int_{t''}^{t'}d\tau \lsb \tilde{\ene}_{a}(\tau) -i\Delta(\tau)\rsb \rcb \approx
\exp\lcb-i\lsb \tilde{\ene}_{a}(t') -i\Delta(t')\rsb(t'-t'') \rcb
\end{equation} 
and replace $t''$ with $t'$ in 
the second term of \eqref{eq:eom_c1}, while the last term vanishes\cite{Brako1980}.
The nonzero terms may be written as
\begin{equation}
	\begin{aligned}
		e^{i\ene_{k}t}c_{k}(t)&=\sum_{k'}e^{i\ene_{k'}t_{0}}c_{k'}(t_{0})\lsb \delta_{k,k'} -i\int_{t_{0}}^{t}dt' \
	e^{i\lb\ene_{k}-\ene_{k'}\rb t'}\tilde{V}^{*}_{ak}(t')\tilde{V}_{ak'}(t') \right. \\
	&\left. \times \int_{t_{0}}^{t'}dt''\exp\lcb-i\lsb \tilde{\ene}_{a}(t') -i\Delta(t')\rsb \rcb (t'-t'') \rsb.
\end{aligned}
	\label{eq:eom_c2}
\end{equation}
The time integral inside the exponential may be performed explicitly giving
\begin{equation}
	\begin{aligned}
		e^{i\ene_{k}t}c_{k}(t)&=\sum_{k'}\int_{t_{0}}^{t}dt'\frac{e^{i\lb\ene_{k}-\ene_{k'}\rb t'}}{2\pi\rho(\ene_{k'})} \
		\lsb 1 -\frac{\Delta(t')}{\Delta(t')-i \lsb \ene_{k'} - \tilde{\ene}_{a}(t') \rsb} \rsb \
		e^{i\ene_{k}t_{0}}c_{k}(t_{0}).
       \end{aligned}
	\label{eq:eom_c3}
\end{equation}
Introducing the phase shift
\begin{equation}
	\begin{aligned}
		\delta_{k}(t)=\arctan\lsb \frac{\Delta(t)}{\ene_{k}-\tilde{\ene}_{a}(t)}\rsb,
       \end{aligned}
	\label{eq:phaseshift}
\end{equation}
\eqref{eq:eom_c3} further simplifies to
\begin{equation}
	\begin{aligned}
		e^{i\ene_{k}t}c_{k}(t)&=\sum_{k'}\int_{t_{0}}^{t}dt'\frac{e^{i\lb\ene_{k}-\ene_{k'}\rb t'}}{2\pi\rho(\ene_{k'})} \
		e^{-2i\delta_{k'}(t')}\
		e^{i\ene_{k}t_{0}}c_{k}(t_{0}).
       \end{aligned}
	\label{eq:eom_c4}
\end{equation}
The slow motion limit also implies that the time scale $\tau$ of the perturbation 
is large, meaning that only $\ene_{k}\approx \ene_{k'}$ contribute
to the integration above. In other words, $\ene_{k}-\ene_{k'}\approx \frac{\hbar}{\tau}$.
It is therefore reasonable to write $\ene_{k}$ for $\ene_{k'}$ in $\delta_{k'}(t')$
and take $\rho(\ene_{k})\approx\rho(\ene_{k'})$.
Taking into account these considerations, we insert $c_{k}(t)$ 
and its corresponding conjugate into \eqref{eq:Hm0} giving
\begin{equation}
	H_{M} = \sum_{k}\frac{\ene_{k}}{4\pi^{2}\rho^{2}(\ene_{k})}\sum_{k'}\sum_{k''} \
	\int_{-\infty}^{\infty}dt\int_{-\infty}^{\infty}ds e^{i\ene_{k}\lb t-s\rb} e^{i\ene_{k''}s}e^{-i\ene_{k'}t}\
	e^{-2i\lsb \delta_{k}(t)-\delta_{k}(s) \rsb}c^{\dagger}_{k''}(t_{0})c_{k'}(t_{0}),
	\label{eq:Hm1}
\end{equation}
where we have formally introduced the integration limits $t=\infty$ and $t_{0}=-\infty$.
Performing the $k$ summation and lifting the time-dependence of the operators results in
\begin{equation}
	\begin{aligned}
		H_{M} &= \lsb 2i\pi\rho(\ene_{k})\rsb^{-1}\sum_{k'}\sum_{k''} \
	\int_{-\infty}^{\infty}dt\int_{-\infty}^{\infty}ds \delta'(t-s)e^{i\ene_{k''}s}e^{-i\ene_{k'}t}\
	e^{-2i\lsb \delta(\ene_{k},t)-\delta(\ene_{k},s) \rsb}c^{\dagger}_{k''}c_{k'}\\
	&=-\lsb 2i\pi\rho(\ene_{k})\rsb^{-1}\sum_{k'}\sum_{k''} \
	\int_{-\infty}^{\infty}dt\int_{-\infty}^{\infty}ds \delta(t-s)\frac{dF_{k'',k'}(t,s)}{dt}\
	c^{\dagger}_{k''}c_{k'},
\end{aligned}
	\label{eq:Hm2}
\end{equation}
where $\delta'(t-s)$ is the derivative of the Dirac delta function and
\[
	F_{k'',k'}(t,s)=e^{i\ene_{k''}s}e^{-i\ene_{k'}t}
e^{-2i\lsb \delta_{k}(t)-\delta_{k}(s) \rsb}.
\]
One of the time integrals may be evaluated to give 
\begin{equation}
	\begin{aligned}
		H_{M} &= \sum_{k'}\sum_{k''}\lsb\frac{\ene_{k'}}{2\pi\rho(\ene_{k})} \
		\int_{-\infty}^{\infty}dte^{i(\ene_{k''}-\ene_{k'})t}c^{\dagger}_{k''}c_{k'}\
		+\frac{1}{\pi\rho(\ene_{k})}\int_{-\infty}^{\infty}dt\dot{\delta}_{k}(t)\
		e^{i(\ene_{k''}-\ene_{k'})t}c^{\dagger}_{k''}c_{k'}\rsb,
\end{aligned}
	\label{eq:Hm3}
\end{equation}
which can be expressed in the desired form as
\begin{equation}
	\begin{aligned}
		H_{eh} \equiv H_{M} = \sum_{k}\ene_{k}\cc\ca
		+\sum_{k}\sum_{k'}W_{k,k'}c^{\dagger}_{k}c_{k'},
\end{aligned}
	\label{eq:Heh}
\end{equation}
where
\begin{equation}
		W_{k,k'}= \frac{1}{\pi\rho(\ene_{k})}\int_{-\infty}^{\infty}dt\dot{\delta}_{k}(t)\
		e^{i(\ene_{k}-\ene_{k'})t}.
	\label{eq:Wk}
\end{equation}
We bosonize \eqref{eq:Heh} using the standard (Tomonaga or Lutinger bosonizations) 
methods\cite{Mahan2000}.
We begin by recognizing that only transitions from occupied to unnoccupied
states are allowed. Hence, since $\ene_{k}\approx\ene_{k'}$, we are therefore restricted
to states close to the Fermi level $\ene_{F}$. 
We may approximate $\dot{\delta}_{k}(t)=\dot{\delta}(v_{F}k,t)\approx\dot{\delta}(\ene_{F},t)$.
Let us denote $L$ as the length of 
our effectively one-dimensional system and let $\ene_{k}=v_{F}k$ and 
$\rho(\ene_{k})\approx\rho(\ene_{F})=\frac{L}{\pi v_{F}}$ giving
\begin{equation}
	\begin{aligned}
		H_{eh} = \sum_{k}v_{F}k\cc\ca
		+\frac{1}{\pi \rho(\ene_{F})}\sum_{k,p}\int dt \dot{\delta}(\ene_{F},t)\exp(iv_{F}pt)c^{\dagger}_{k}c_{k-p},
\end{aligned}
	\label{eq:Heh1}
\end{equation}
By defining the coherent e-h pairs boson operators 
\[
	b_{q}=\lb\frac{2\pi}{qL} \rb^{-1/2}\sum_{k}c^{\dagger}_{k}c_{k-q}
\]
\[
	b_{q}^{\dagger}=\lb\frac{2\pi}{qL} \rb^{-1/2}\sum_{k}c^{\dagger}_{k-q}c_{k},
\]
\eqref{eq:Heh1} may be
expressed in the boson representation as
\begin{equation}
	\begin{aligned}
		H_{eh} = \sum_{q}\omega_{q}\bc\ba + \sum_{q}\lsb f_{q}(x)\ba + f_{q}^{*}(x)\bc \rsb,
\end{aligned}
	\label{eq:Hehbos}
\end{equation}
where $\omega_{q} = v_{F}q$ and  
\begin{equation}
	f_{q} = \lsb\frac{2\omega_{q}}{\pi^{2} \rho(\ene_{F})}\rsb^{1/2} \int dt \dot{\delta}(\ene_{F},x(t))\exp(i\omega_{q}t).
	\label{eq:fq}
\end{equation}
The time-dependence of the phase shift is formally indicated
through the particle coordinate $x(t)$ ($R(t)$ in the main text) 
by  $\delta(t)\rightarrow \delta(x(t))$.

\section{Frictional Force Kernels}
\label{supp:elec_fric}
The frictional force term of \eqref{eq:Langevin_1} may be written as
\begin{equation}
	\int ds\lsb \Gamma_{1}(t-s)+ \Gamma_{2}(t-s)\rsb\dot{R}(s)
	\label{eq:fric_ker}
\end{equation}
where
\begin{equation}
	\int ds \Gamma_{1}(t-s)\dot{R}(s)=\frac{1}{2}\int ds \sum_{j}\frac{c'_{j}(R(t))c_{j}^{\prime *}(R(s))}{\Omega_{j}}\
	[\mathrm{sgn}(t-s)+1]\cos\Omega_{j}(t-s)\dot{R}(s),
\label{eq:fric_kerR1}
\end{equation}
and
\begin{equation}
	\int ds \Gamma_{2}(t-s)\dot{R}(s)=\frac{1}{2}\int ds \sum_{q}\frac{f'_{q}(R(t))f_{q}^{\prime *}(R(s))}{\omega_{q}}\
	[\mathrm{sgn}(t-s)+1]\cos\omega_{1}(t-s)\dot{R}(s).
\label{eq:fric_kerR2}
\end{equation}
We may express \eqref{eq:fric_kerR1} as
\begin{equation}
	\int ds \Gamma_{1}(t-s)\dot{R}(s)=\frac{1}{\pi}\int ds \
	\int d\Omega \frac{\mathscr{J}(\Omega)}{\Omega}\dot{R}(s)\
	[\mathrm{sgn}(t-s)+1]\cos\Omega(t-s),
\label{eq:fric_kerR1_1}
\end{equation}
where the spectral density $\mathscr{J}(\Omega)$ is defined as $\mathscr{J}(\Omega)=\
\pi\sum_{j}\frac{c'_{j}(R(t))c_{j}^{\prime *}(R(s))}{2}\delta(\Omega-\Omega_{j})$.
We assume the particle coupling with the bosonic bath is linear $c_{j}(R)=Rc_{j}$,
so that $\mathscr{J}(\Omega)=\
\pi\sum_{j}\frac{c_{j}c_{j}^{*}}{2}\delta(\Omega-\Omega_{j})$. 
We choose $\mathscr{J}(\Omega)=\gamma\Omega$, where $\gamma$ is the bosonic bath
coefficient of friction, \eqref{eq:fric_kerR1_1} is simplified as
\begin{equation}
	\int ds \Gamma_{1}(t-s)\dot{R}(s)=\gamma\dot{R}(t)\
\label{eq:fric_kerR1_2}
\end{equation}
For the electronic frictional force, we insert \eqref{eq:fq} ($x(t)\to R(t)$) 
into \eqref{eq:fric_kerR2},
and change the sum over $q$ to an integral over $\omega$ yielding
\begin{equation}
	\int ds \Gamma_{2}(t-s)\dot{R}(s)=\int ds \int d\omega\
	\delta(\omega)\delta(-\omega)\
	[\mathrm{sgn}(t-s)+1]\cos\omega(t-s)\dot{R}(s),
\label{eq:fric_kerR2_1}
\end{equation}
where
\begin{equation}
	\delta(\omega)=\frac{1}{\pi}\int dt \dot{\delta}(\ene_{F},R(t))e^{i\omega t},
\label{eq:ps_ft}
\end{equation}
with $\delta(\ene_{F},R(t))$ being the phase shift (\eqref{eq:phaseshift}).
The integrand in \eqref{eq:fric_kerR2_1} is peaked at $t=s$ and
can be replaced by a delta function of the same area\cite{Brandbyge1995,Newns1985}
giving
\begin{equation}
	\begin{aligned}
	\int ds \Gamma_{2}(t-s)\dot{R}(s)&=\Bigg|\frac{d\delta(\ene_{F},R)}{dR} \Bigg|^{2}\dot{R}(t),\\
	&=\eta(R)\dot{R}(t),
\end{aligned}
\label{eq:fric_kerR2_2}
\end{equation}
where the electronic friction coefficient $\eta(R)$ is given by 
\begin{equation}
	\eta(R)=\pi\lcb\frac{d\Delta(R)}{dR}\frac{\lsb \ene_{F}-\tilde{\ene}_{a}(R)\rsb}{\Delta(R)}\
	+\frac{d\tilde{\ene}_{a}(R)}{dR}\rcb^{2}\rho_{a}^{2}(\ene_{F},R),
	\label{eq:sfric}
\end{equation}
(\eqref{eq:fric} in the main text).

\section{Simulation Parameters}

In both numerically worked out problems, we set $m=1$ and $\hbar=1$ in the Langevin simulations.
In the solvated proton discharge problem, we represent
the resonance width and the proton energy levels as Gaussian functions
of $R$, $\Delta(R)=\Delta_{0}\exp[-\alpha(R-R_{0})^{2}]$ and
$\ene_{a}(R)=(\ene_{0}-\ene_{\infty})\exp[-\kappa(R-R_{0})^{2}]+\ene_{\infty}$ 
where $\Delta_{0}=0.01$ eV, $\alpha=0.015~a_{0}^{-2}$, $\kappa=0.05~a_{0}^{-2}$
$\ene_{\infty}=\ene_{F}=0.0$ eV, $\ene_{0}=-1.0$ eV, and $R_{0}=2.0~a_{0}$.
Here, $\alpha$ is proportional to the surface wavefunction decay into the solvent
and $\kappa$ is chosen in such a way that the energy level crosses $\ene_{F}$
near the barrier of the following excited state potential energy surface (PES)
\begin{equation}
	V_{H^{+}}(R)=V(R)+\tilde{\ene}_{a}(R).
\label{eq:ex_pes}
\end{equation}
Here, $V(R)=A\lcb \lsb s(R-\sigma)\rsb^{4} - \lsb s(R-\sigma)\rsb^{2} + a(R-\sigma)\rcb$,
where $A=1.5$ eV, $s=0.25~a_{0}^{-1}$, $\sigma=5.0~a_{0}$ and 
$a=-0.05~a_{0}^{-1}$ and $\tilde{\ene}_{a}(R)$ is the proton energy level.
\begin{figure}
  \centering
	  \includegraphics[scale=0.72]{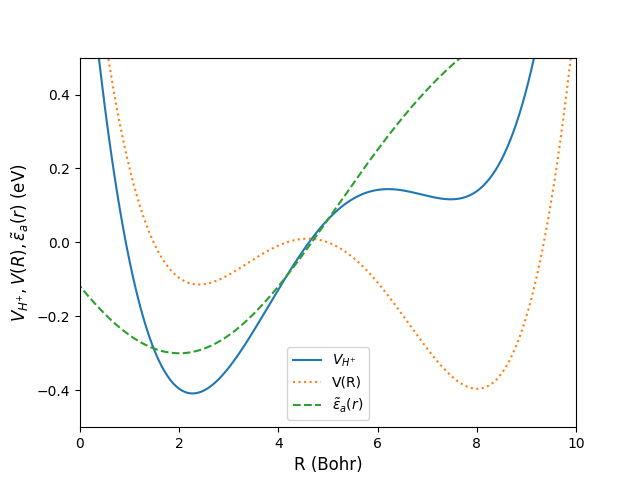}
  \captionsetup{justification=raggedright,singlelinecheck=false}
  \caption{Plots of $V_{H^{+}}(R)$, $V(R)$ and $\tilde{\ene}_{a}(R)$
  }
  \label{fig:ex_pes}
\end{figure}
\figref{fig:ex_pes} shows the plots of $V_{H^{+}}(R)$, $V(R)$ and $\tilde{\ene}_{a}(R)$.
$V(R)$ is an assymetric double well potential with a deeper well at
$R_{m1}=8.0~a_{0}$ and a shallower well at $R_{m2}=2.0~a_{0}$.
This shape is typical of a Volmer process obtained from ground state ab initio calculations.
As we are dealing with a charged particle, we consider the excited state PES of
the proton described by $V_{H^{+}}(R)=V(R)+\tilde{\ene}_{a}(R)$ instead of $V(R)$. 
The addition of $\tilde{\ene}_{a}(R)$
causes the substantial deepening and shallowing of the PES wells at $R_{m2}$ and $R_{m1}$,
respectively. 
As we have seen in the formulations, the PES should be renormalized by the couplings to the solvent
baths and the electronic system. We have absorbed these renormalizations in the definitions of
the parameters above to simplify the simulations. 


\bibliography{Ref_PI.bib}

\makeatletter\@input{xx.tex}\makeatother